%% file: TDTonPBF_0_main.tex
\documentclass[conference]{IEEEtran}
\ifCLASSINFOpdf
\else
\fi

\input{TDTonPBF_1_Packages}
\input{TDTonPBF_2_Commands}

\begin{document}
%

\input{TDTonPBF_3_Title}

\input{TDTonPBF_4_Authors}



\maketitle


\input{TDTonPBF_5_Abstract}


%
\IEEEpeerreviewmaketitle

\input{TDTonPBF_7_Sections}

\bibliography{TDTonPBF_references}{}
\bibliographystyle{IEEEtran}
%



\balance

\input{TDTonPBF_8_Appendix}

\end{document}

%% file: TDTonPBF_1_Packages.tex
\usepackage{booktabs} 
\usepackage{listings}
\usepackage{paralist}

\usepackage{incgraph,tikz}
\usepackage{graphicx}
\usepackage{longtable}
\usepackage{supertabular}

\usepackage{wrapfig}
\usepackage{subcaption}
\usepackage{hyperref}
\usepackage{multirow} 
\usepackage{balance}

\usepackage{enumitem}
\usepackage{csquotes}

\usepackage{multirow}
\usepackage{pifont}
\usepackage{xcolor,colortbl}

\usetikzlibrary{positioning,shapes.geometric}

\usepackage{soul}

\usepackage{enumitem}

\usepackage{comment}
\usepackage{amsmath}

\usepackage{tikz}
\usetikzlibrary{arrows,shapes,automata,petri,positioning,calc}
\tikzset{
    place/.style={
        circle,
        thick,
        draw=black,
        fill=gray!50,
        minimum size=6mm,
    },
        state/.style={
        circle,
        thick,
        draw=blue!75,
        fill=blue!20,
        minimum size=6mm,
    },
}

%% file: TDTonPBF_2_Commands.tex
\newcommand{\todo}[1]{\textcolor{red}{\textbf{TODO: #1}}}
\newcommand{\ask}[2]{\textcolor{red}{\textbf{@{#1}: #2}}}

\newcommand{\rev}[1]{\textcolor{red}{#1}}
\newcommand{\revnr}[1]{\textcolor{red}{#1}}
\newcommand{\remopt}[1]{\textcolor{magenta}{#1}} 
\newcommand{\remfirst}[1]{\textcolor{magenta}{#1}} 
\newcommand{\remsecond}[1]{\textcolor{magenta}{#1}} 

\renewcommand{\todo}[1]{}
\renewcommand{\ask}[2]{}
\renewcommand{\rev}[1]{#1}
\renewcommand{\revnr}[1]{#1}
\renewcommand{\remopt}[1]{#1}
\renewcommand{\remfirst}[1]{}
\renewcommand{\remsecond}[1]{#1}

%% file: TDTonPBF_3_Title.tex
\title{Turning Hearsay into Discovery: \\ Industrial 3D Printer Side-Channel Information Translated to Stealing the Object Design}

%% file: TDTonPBF_4_Authors.tex
\author{\IEEEauthorblockN{Aleksandr Dolgavin$^{1}$,
Jacob Gatlin$^{1}$,
Moti Yung$^{2, 3}$, and
Mark Yampolskiy$^{1}$}
\IEEEauthorblockA{$^{1}$Auburn University,~ $^{2}$Google LLC,~ $^{3}$Columbia University}
}







%% file: TDTonPBF_5_Abstract.tex
\begin{abstract}

The central security issue of outsourced industrial 3D printing (aka AM: Additive Manufacturing), a critical industry that is expected to dominate manufacturing, is the protection of the digital design (containing the designers' model, which is their intellectual/ confidential property) shared with the manufacturer.
Here, we show, for the first time,  that side-channel attacks are, in fact, a concrete serious threat to existing industrial-grade 3D printers, enabling the reconstruction of the model printed (regardless of employing ways to directly conceal the design, e.g. by encrypting it in transit and before loading it into the printer).
Previously, such attacks were demonstrated only on fairly simple FDM desktop 3D printers which play a negligible role in manufacturing of valuable designs.

We focus on the Powder Bed Fusion (PBF) AM process, which is popular for manufacturing net-shaped parts with both polymers and metals.
We demonstrate how its individual actuators can be instrumented for the collection of power side-channel information during the printing process.
We then present our approach to reconstruct the 3D-printed model solely from the collected power side channel data.
Further, inspired by Differential Power Analysis, we developed a method to improve the quality of the reconstruction based on multiple traces.
We tested our approach on two design models with different degrees of complexity.
For different models, we achieved as high as 90.29~\% of True Positives and as low as 7.02~\% and 9.71~\% of False Positives and False Negatives by voxel-based volumetric comparison between reconstructed and original designs.

The lesson learned from our attack is that the security of AM design files cannot rely solely on protecting the files themselves in an industrial environment. 
Instead, it must also rely on ensuring that no leakage of power, noise, and similar signals can be detected by potential eavesdroppers in the printer's vicinity.
 
\end{abstract}

%% file: TDTonPBF_7_Sections.tex

\section{Introduction}

\emph{Additive Manufacturing} (AM), commonly known as \emph{3D Printing}, is a critical manufacturing technology that produces physical objects incrementally, typically adding and fusing thin layers of source material~\cite{astmF2792}. 
AM is used to manufacture parts with a wide range of materials, from soft polymers (plastics) to high-performance metal alloys, and even biological cells. 
Due to numerous technical and economic advantages, this technology has experienced rapid adoption in a wide range of applications, including rapid prototyping, medical implants, and components of safety-critical systems such as jet or rocket engines. 
By 2024, the total size of the rapidly expanding AM market is estimated at \$21.9 B and is expected to increase to \$114.5 B in the next decade~\cite{wohlers2025report}.

Although manufacturing ``in house'' is often a preferred solution by large companies like General Electric (GE), the required capital investment is, in fact, prohibitive for smaller players: 
industrial-grade polymer machines cost tens to hundreds of thousands, and for metal machines, from multiple hundreds of thousands to several million dollars~\cite{wohlers2025report}.
In addition, because AM requires literally no retooling and always operates on the basis of a digital design file, it offers an increasingly popular
cost-saving outsourcing solution. That is, the actual production of parts (in small or large quantities) can be outsourced to an AM company that offers manufacturing-as-a-service. 
This business model is also beneficial to AM service providers, who can achieve high utilization of their machines, and thus better Return on Investment (ROI).
Currently, it is not surprising that this business model is popular with both service providers and customers~\cite{yampolskiy2022state}.
This is reflected in the fact that more than \revnr{2,000} companies offer AM manufacturing service~\cite{url20243dprintingbusiness}.

At the same time, the outsourcing business model opens the door to concerns about the appropriate handling of digital design files shared with the manufacturer~\cite{yampolskiy2022state}.
To address this concern, several solutions have been proposed by both academia and industry.
The academic proposal to prevent design theft included various Digital Rights Management (DRM) architectures~\cite{wade2016digital, alkaabi2020blockchain}, 
direct streaming to the 3D printer~\cite{baumann2017model, kolter2025streaming, kok2022design}, and the integration of a Trusted Platform Module (TPM) on the machine to allow its secure boot and end-to-end encryption~\cite{safford2019hardware, cultice2023novel}.
Among the commercial solutions, the most notable are developed by Identify3D~\cite{identify2020info} (currently part of Materialise~\cite{materialise2022pressrelease}) and Assembrix~\cite{assembrix2023web}. 
Identify3D offers AM design-centric access control on the manufacturer site, while Assembrix implements cloud-based design handling and streaming directly to the used 3D printer.

However, in the case of a malicious manufacturer or a malicious insider, the designs are exposed to the printer owner in a threat model known as Man-at-the-End (MATE)~\cite{collberg2009surreptitious, akhunzada2015man}.
Note that protecting against MATE is very challenging and, in many cases, infeasible.
When it comes to desktop 3D printers, 
several researchers first warned~\cite{yampolskiy2014intellectual} and then demonstrated~\cite{faruque2016acoustic, hojjati2016leave, song2016my, gatlin2021encryption, pearce2022flaw3d} that side-channel attacks can be used to obtain digital designs.
Although the original attack showed reconstruction results of just \revnr{78.35~\%} accuracy in axis prediction and an average error of \revnr{17.82~\%} in estimating the movement distance for a simple key profile based on acoustic emanations~\cite{faruque2016acoustic}, a more recent attack based on the power side-channel achieved a spatial reconstruction of more than \revnr{99~\%} with ten models of varying complexity~\cite{gatlin2021encryption}. 
Regarding the implications to the outsourcing AM industry, though,
to the best of our knowledge, all side-channel attacks have been demonstrated only against Fused Deposition Modeling (FDM), a technology very popular with inexpensive desktop 3D printers, but playing a negligible role in the outsourced manufacturing of valuable designs.

Therefore, to date, the ``(much beyond) million-dollar question'' remained whether significantly more complex industrial-grade AM machines are also susceptible to side-channel attacks.
In this paper, we provide a positive answer to this question,
specifically for the Powder Bed Fusion (PBF) AM machines, which is the dominant technology for manufacturing net-shaped parts with both polymers and metals.
To the best of our knowledge, this is the first time that side-channel attacks stealing designs (for espionage or I.P. stealing purposes) have been proven to be possible on industrial-grade 3D printers.
Hence, the attacks we present are a real warning to the AM industrial ecosystem's players.


\section{Related Work}
\label{sec:related_work}

\emph{Side-channels} are (unintended) physical emanations produced by cyber or cyber-physical processes. 
To name just a few, they include power draw~\cite{randolph2020power}, electromagnetic (EM) emanations~\cite{sayakkara2019survey}, and execution timing~\cite{ge2018survey}.
Whenever there is a tight correlation between the measured physical signal's property and the process that caused this emanation, it becomes possible to reason about this process by analyzing the side-channel data.

Side-channels have been studied and exploited in the context of various systems, most famously for attacks, but also for defense measures.
Such attacks have been shown to be a very effective means of recovering cryptographic keys used by an embedded system (ES)~\cite{kocher1999differential, ravi2004security} or smart cards\cite{mangard2007power, rankl2003overview, standaert2009unified}.
Similar attacks have also been shown in the context of IoT devices~\cite{devi2020side, zankl2018side}.
In the context of smart grids, they can be used to violate user privacy, exploiting unique power usage patterns of appliances~\cite{quinn2009privacy}.
In the context of cloud systems, they have been shown to be capable of reasoning about another (cyber-separated) virtual machine (VM) running on the same hardware~\cite{ristenpart2009hey, zhang2012cross}.
These often exploit CPU or cache implementation to bypass address space isolation, as has been demonstrated by attacks such as Meltdown~\cite{lipp2018meltdown} or Spectre~\cite{kocher2020spectre, koruyeh2018spectre}.
In fairness, there are also multiple examples of benevolent use of side-channels, such as otherwise difficult-to-achieve detection of hardware Trojans~\cite{agrawal2007trojan, narasimhan2012hardware, tehranipoor2010survey}.

\begin{table}[tbp]
  \centering
  \begin{tabular}{lll}
    \toprule
    \textsc{3D Pr.} & \textsc{Side-Channel(s)} & \textsc{Publication}  \\
    \midrule
    FDM & Acoustic & Al Faruque et al.~\cite{faruque2016acoustic}  \\


     FDM &  Acoustic, Inertial & Song et al.~\cite{song2016my} \\

     FDM &  Acoustic, Magnetic & Hojjati et al.~\cite{hojjati2016leave} \\

     FDM & Inertial, Magnetic, Optical &  Gao et al.~\cite{gao2018watching} \\ 

     FDM &  Power & Gatlin et al.~\cite{gatlin2021encryption} \\
     
     FDM & Vibration &   Stańczak et al.~\cite{stanczak2021vibration} \\
     
     FDM & Digital signal (chip insert) & \remopt{Pearce et al.~\cite{pearce2022flaw3d}} \\ 
          
     FDM & Acoustic, Magnetic &  Jamarani et al.~\cite{jamarani2024practitioner} \\

     FDM & Optical &   \remopt{Chattopadhyay et al.~\cite{chattopadhyay2025one}} \\
     
    \bottomrule
  \end{tabular}
  \caption{Works on side-channel-based reconstruction of 3D-printed designs. These were attacks against fairly simple Fused Deposition Modeling (FDM) {\bf desktop 3D printers}. \todo{?? add Spatial-temporal}}
  \label{tab:side-channel-TDTattacks}
\end{table}

In the AM context, side-channels can be used to reconstruct the design of the 3D-printed part.
After the possibility of such attacks was raised by Yampolskiy et al.~\cite{yampolskiy2014intellectual}, 
the first practical demonstration of it was provided by Al Faruque et al.~\cite{faruque2016acoustic}.
In a Fused Deposition Modeling (FDM) 3D printer, movement of the extrusion head along all the three axes and the extrusion of the heated filament are performed by stepper motors.
The researchers recognized that the acoustic emanations generated by these motors depend on the direction and speed of movement. 
As it turned out, the acoustic emanations generated by different motors were distinguishable, enabling recognition with a certain probability along which axis the movement was conducted.
The researchers developed a ``machine learning'' solution that achieved \revnr{78.35~\%} accuracy in axis prediction, giving \revnr{17.82~\% }length error when tested against 3D printing of an outline of a key. 
Thereafter, several other works followed, showing reconstruction of the design based on side-channel information.
They differed in the side-channel(s) used, the way the side-channel data was collected, the method applied to reconstruct the design, how the reconstruction quality was tested, and how the reconstruction results were reported.

Of all these works, the most relevant to our research is the publication by Gatlin et al.~\cite{gatlin2021encryption} who
instrumented the power side-channel of individual actuators (on the used 3D printer, all of them were stepper motors).
The authors then used their understanding of how stepper motors are driven by the supplied current to reconstruct the printed object with impressive spatial accuracy \revnr{99~\%}.

Here, however, the technical similarity to our work ends. 
Although we reuse the general concept of instrumenting individual actuators and measuring the power side-channel, the actuators we used in the reconstruction are entirely different and we had to {\em``reverse engineer'' the correlation between the measured signals and the action taken}.  Namely, we use a step which can be called \emph{``known design attacks''} 
(akin to ``known plaintext attacks'' on cryptosystems, and inspired by ``template attacks''~\cite{batina2019online})
to learn how to interpret/ correlate the printer's signal.
This stronger attack (which is nevertheless doable by insiders who activate/ maintain/ operate/ test the printer) is the key to threats to the more complicated industrial printers.
Note that the common theme across all similar prior attacks was that {\em they have all been demonstrated against simple non-industrial FDM 3D printers} (see the summary in Table~\ref{tab:side-channel-TDTattacks}).


\section{Background: Powder Bed Fusion (PBF)}
\label{sec:background:pbf}

\emph{Powder Bed Fusion} (PBF) is one of seven distinct AM processes~\cite{astmF2792}, each of which has further refinements, often referred to as AM technologies.
Its ability to operate with a wide range of materials (from polymers to high-performance metal alloys) and unrivaled geometric precision of manufactured parts made PBF one of the most dominant technologies in industrial settings. 
Although there are differences between PBF machines that work with metals and polymers, they are related to the material handling and, as such, are of no relevance to this work.
In what follows, we focus on PBF that uses a laser as a heat source.

\begin{figure}[tbp]
  \centering
  \includegraphics[width=0.9\linewidth]{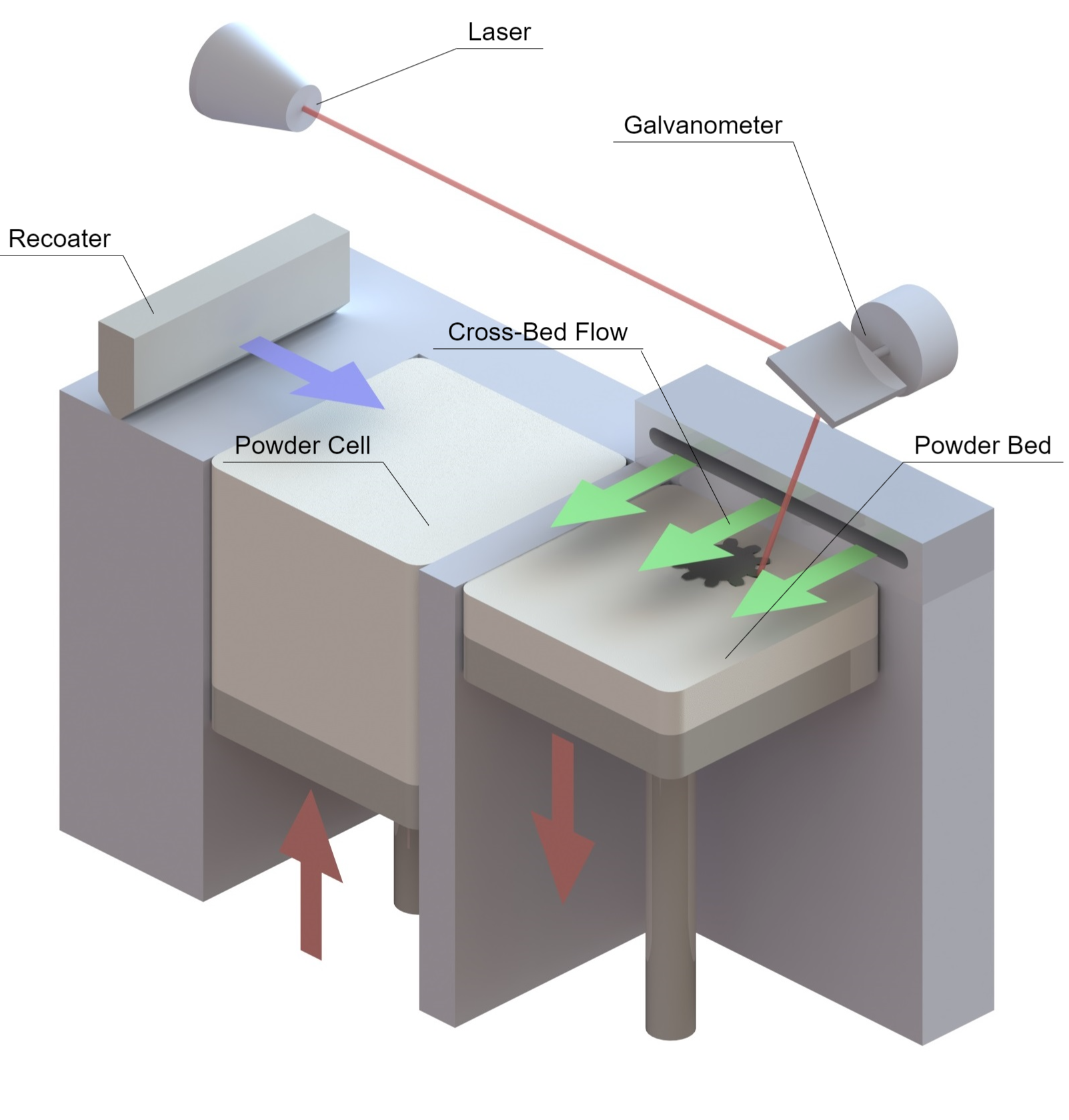}
  \caption{Schematic of a Laser Beam PBF machine \cite{zinner2022spooky}.}
  \label{fig:lbpbf}
\end{figure}

A general concept of how a PBF machine works is depicted in Figure~\ref{fig:lbpbf}.
The machine utilizes the source material \remopt{(a.k.a. \emph{feedstock})} in powder form. 
The process alternates between the powder distribution and the layer sintering phases.
 
In the powder distribution phase (or, for our purposes, \emph{layer transition}), a thin layer of source material is evenly distributed on top of \emph{powder bed}.
For this, the platform under it is lowered to create a vertical opening according to the intended layer height.
The \emph{powder cell} that contains the unmelted powder is raised, to expose enough powder to completely cover this opening. 
Afterwards, the \emph{recoater} blade pushes and evenly distributes the exposed fresh powder from the powder cell across the powder bed. 
\remfirst{Upon return of the recoater to the initial position, the part profile can be sintered or melted using a heat source.}

In the \emph{layer sintering} phase, the laser sinters the profile that corresponds to the specific layer of the part, which is also fused to the underlying layer.
As the laser source is stationary, the laser beam is deflected along the X and Y axes by two \emph{galvanometer} units (in the figure, only one is depicted), each of which is basically a fast-moving mirror whose rotational angle can be controlled precisely. 
Afterwards, the machine switches back to the layer transition phase, and the process repeats until all layers of the part are ``printed''.


\section{Instrumentation \& Experimentation}
\label{sec:instrumentation_experimentation}

\subsection{Used PBF 3D Printers: Sintratec S2}

For our work, we decided to use a polymer PBF machine.
Two factors played the most significant role in our decision.
First, the type of actuators and the ways how they utilized is shared by both polymer and metal PBF machines (described in Section~\ref{sec:background:pbf}).
Second, metal powders are hazardous and can be combustible~\cite{vander2004laboratory}, which already led to severe injuries~\cite{osha2014after}.
\remfirst{In addition, capital investment and operation costs of metal AM is by at least one order of magnitude higher than of a polymer one.}

Upon evaluating the options available, we selected the Sintratec S2. 
With a few minor differences (such as having two bins of fresh powder), its inner working follows the general concept described in Section~\ref{sec:background:pbf}.
The only other difference that will play a role in the later discussion is that this machine uses UV lamps to preheat the build chamber (and thus the powder) to a set temperature.
We will return to the implications of this in Section~\ref{sec:reconstruction:initial_analysis}, where we decide about the information from which actuator should be used for object reconstruction.

\subsection{Instrumentation \& Data Acquisition}
\label{sec:instrumentation}

\begin{figure}[tbp]
    \centering

    \begin{subfigure}[b]{0.5\linewidth}
        \centering
		\includegraphics[width=\linewidth]{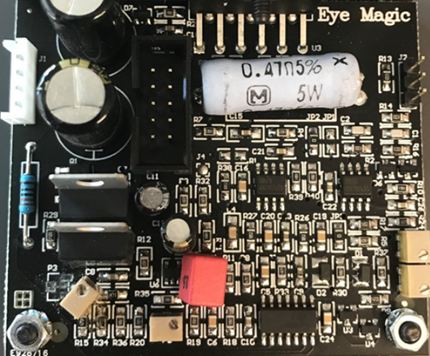}
            \caption{S2 galvanometer control board}
            \label{fig:GalvanometerControlBoard}
    \end{subfigure}%
    ~ ~ ~
    \begin{subfigure}[b]{0.5\linewidth}
        \centering
		\includegraphics[width=\linewidth]{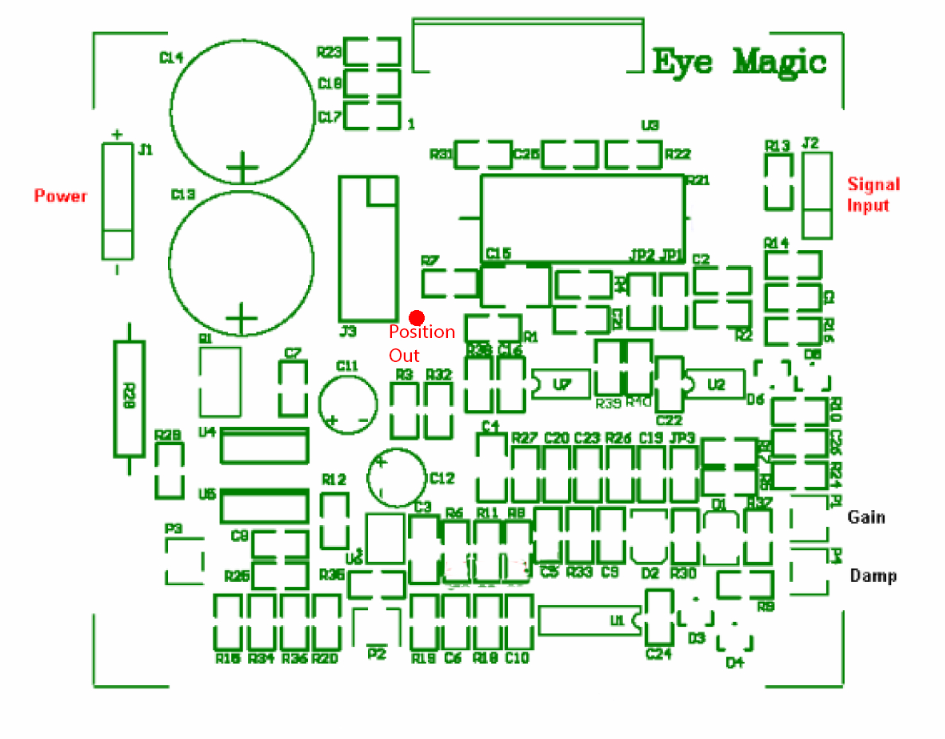}
            \caption{Schematic of Eye Magic}
            \label{fig:galvo_schematic}
    \end{subfigure}%

    ~

    \begin{subfigure}[b]{0.5\linewidth}
        \centering
		\includegraphics[width=\linewidth]{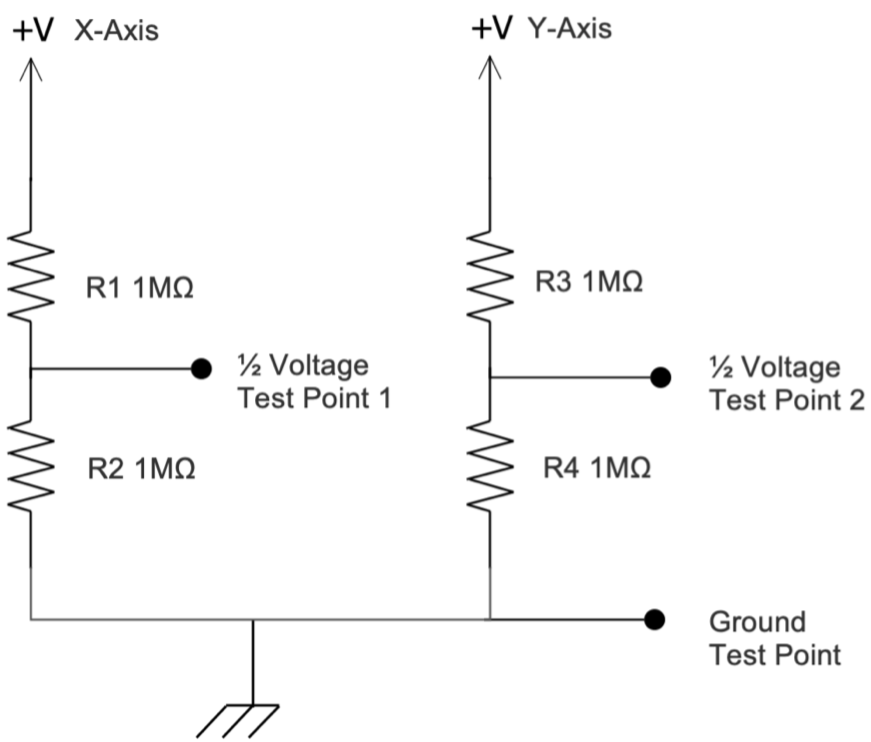}
            \caption{Voltage-divider circuit}
            \label{fig:voltage_divider}
    \end{subfigure}%
    ~ ~ ~    
    \begin{subfigure}[b]{0.5\linewidth}
        \centering
		\includegraphics[width=\linewidth]{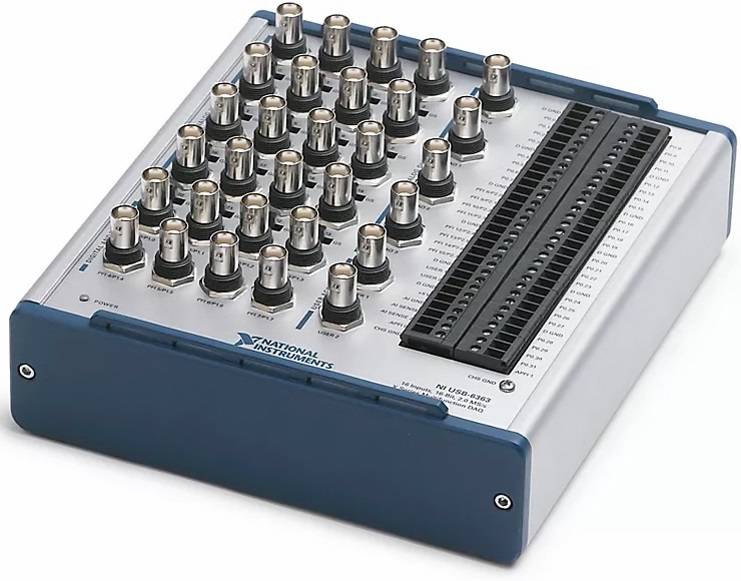}
            \caption{DAQ: NI USB-6363} 
		\label{fig:instrumentation:NI_USB-6363}
    \end{subfigure}%

	\caption{Instrumentation of power side-channel (selected). }
	\label{fig:instrumentation}
\end{figure}

For the proposed approach, we instrument the power side-channels of the individual actuators. 
Specifically, we instrumented the laser, both X- and Y- galvanometers, and all stepper motors responsible for the movement of the print and powder beds as well as of the recoater.
The instrumentation varies for different types of actuator, as we outline below.

To instrument stepper motors, we reused the approach originally proposed by Gatlin et al.~\cite{gatlin2021encryption}: The current in both phases of each stepper motor have been measured using inductive current probes. 
We used Fluke i310~\cite{flukei310s} inductive current probes, which can be clamped around the corresponding wires.
The probes are rated to measure current in the ranges 30 A and 300 A AC or ±45 A and 450 A DC with the resolution of ±50 mA (40 A) to ±100 mA (400 A). 

We reused the same approach to instrument the laser signal by measuring its power supply.
Instrumentation of the galvanometers using the same approach posed the following challenge.
Both galvanometers were connected to the main control board via a ribbon cable.
Upon creating a breakout cable and analyzing the signals captured from individual lines, we identified that these are digital logic signals\remopt{, which are typically designed to draw as little current as possible}.
The analog control signals for these devices are routed internally to the galvanometer chassis, preventing non-intrusive instrumentation with inductive current clamps.
Therefore, we were forced to find an alternative solution for the instrumentation of galvanometers.

The S2 has two Eye Magic galvanometer control boards (see Figure~\ref{fig:GalvanometerControlBoard}), schematic for which is shown in Figure~\ref{fig:galvo_schematic}.
The board schematics contains a test point (marked as a red dot in the middle of the board and labeled ``Position Out'').
According to the board documentation, this test point provides a voltage from -10 V to +10 V that corresponds to the intended motor position and, therefore, with mirror orientation and target laser position on the print bed.
To ensure that the measured voltage is indeed in this range and that the current draw is minimal, we designed and built a voltage divider circuit (the schematic is shown in Figure~\ref{fig:voltage_divider}), through which we measured the outputs at the test points.

\remopt{To ensure safe operation with a machine that contains a laser, we redesigned the top of the S2 enclosure.
It allowed relatively simple access to the probes and instrumented devices when S2 was off, and it could be sealed for experimentation.
We routed the cables of all probes outside the machine through a special opening in the enclosure.}

In total we had 11 probes: 1x for laser power, 2x for X- and Y- galvanometer boards, 6x for two powder cells and a print bed, and 2x for the recoater. 
To synchronize across multiple measured channels and to store high-frequency measurements collected during potentially long-duration prints, we used a dedicated 782258-01 National Instruments (NI) Data AcqQisition (DAQ) device~\cite{NIUSB6363} (see Figure~\ref{fig:instrumentation:NI_USB-6363}). 
The DAQ model used has BNC termination for up to 16 analog input channels\remfirst{~and 4 analog output channels}, which was sufficient for our purposes.
The DAQ is monitored and controlled using NI Flexlogger~\cite{NIflexlogger}, generating long-term logs for analysis using NI DIADEM~\cite{NIdiadem}. 

The data acquisition rate must be above the signal-specific threshold for its reconstruction\footnote{According to the Nyquist–Shannon Sampling Theorem~\cite{nyquist1928certain, shannon1948mathematical}, the sampling rate must be at least twice the frequency of every signal component that must be reconstructed without errors.}.
\remopt{Despite different signal characteristics on monitored channels, for the sake of simplicity of data processing, we decided to use the same sampling rate across all channels.}
Based on our preliminary signal analysis, we decided to sample all instrumented power side-channels at a \revnr{20~kHz} rate to guarantee capture of the full signal resolution. 
Even for the galvanometer signals, which are the fastest changing in the machine, this constituted intentional oversampling; the reason was to address a potential event of unexpected high-frequency signal characteristics that need to be investigated.

To support the data collection of longer captures, the DAQ automatically splits the captured data into multiple files, each up to 2GB.
The split files are seamlessly recombined so that there is no loss of data.
All data collected are stored in the proprietary NI TDMS file format. 
Upon completion of the experiments and data collection, we converted TDMS to CSV file format; then we used CSV files as input for the signal processing and object reconstruction software that we developed.

\subsection{Conducted Experiments}
\label{sec:conducted_experiments}

\begin{figure}[tbp]
    \centering
    \begin{subfigure}[b]{0.5\linewidth}
        \centering
		\includegraphics[width=\linewidth]{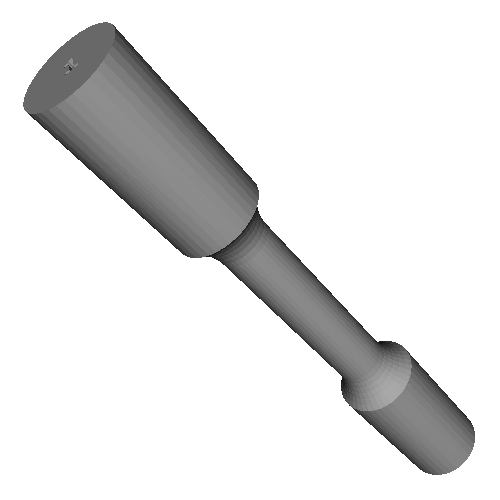}
        \caption{ASTM E8 } 
		\label{fig:test_objects:TestModel_ASTM_E8}
    \end{subfigure}%
    \begin{subfigure}[b]{0.5\linewidth}
        \centering
		\includegraphics[width=\linewidth]{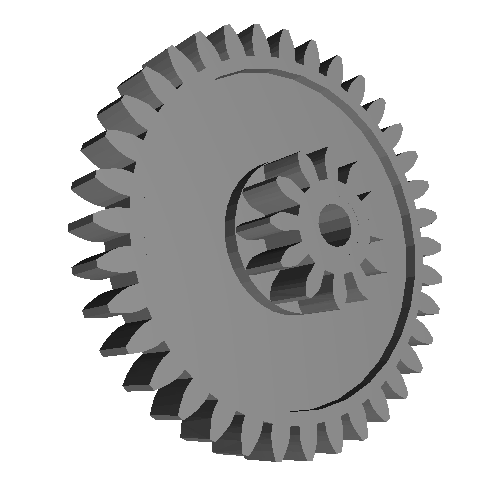}
        \caption{Gear }
		\label{fig:test_objects:TestModel_Gear}
    \end{subfigure}%

	\caption{Models of test objects for which power side-channel traces were collected. 
    Printed: ASTM \revnr{3x} times, Gear \revnr{1x} time. 
    }
	\label{fig:test_objects}
\end{figure}

We selected two designs for experimental test evaluation: an ASTM E8/E8M-22~\cite{ASTME8} test specimen \remopt{(later, for brevity, we will refer to it only as ``ASTM E8'' or simple ``ASTM'')} commonly used in mechanical tests and a Gear model that is representative of more realistic part geometry (see Figure~\ref{fig:test_objects}). 
Both models were specified in STL files. 
\remopt{For printing, they were prepared (\emph{sliced}) using the proprietary Sintratec slicer.}

We printed both models with full (100\%) infill. 
Initially, it was not clear whether power side-channel traces collected during multiple prints of the same object will show substantial deviations, and if it is the case, what the reasons would be.
Specifically, whether there will be significant deviations in (a) signal strength and (b) timing of actuation signals.
To investigate this, we printed the ASTM specimen \revnr{three} times, because the simplicity of its geometry allowed for manual analysis. 
We printed the Gear model only once.


\section{3D-Printed Object Reconstruction}
\label{sec:reconstruction}

\subsection{\remopt{Initial Trace Analysis \& Decisions}}
\label{sec:reconstruction:initial_analysis}

\begin{figure}[tbp!]
    \centering
    \includegraphics[width=.95\linewidth]{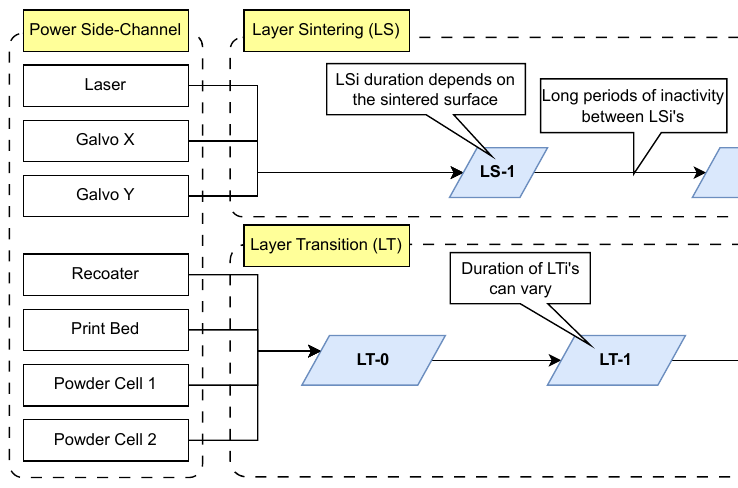}
    \caption{General structure and some fundamental characteristics immediately recognizable from the collected power side-channel traces. Based on initial manual analysis, Laser and Galvanometer X and Y data caries enough information to fully reconstruct 3D-printed object. \todo{?? remove?}}    
    \label{fig:PSCbDetectionSignature}
\end{figure}

To decide on the reconstruction strategy, we manually analyzed the collected traces.
This analysis confirmed our assumption that the side-channel data (measured as described in Section~\ref{sec:instrumentation}) correlate with the actions of individual actuators.
With slight deviations specific to S2, actuator sequencing followed the general operation concept of PBF described in Section~\ref{sec:background:pbf}.
The interchangeable phases of layer transition (LT) and layer sintering (LS), as well as the actuators involved, could be clearly distinguished (see Figure~\ref{fig:PSCbDetectionSignature}).

We then analyzed the durations of LSi's and LTi's on the basis of one of the ASTM specimen traces.
The duration of LSi's was relatively short and ranged between \revnr{3754} and \revnr{26170} data points (\revnr{0.19} to \revnr{1.31~s} at \revnr{20~kHz} data acquisition rate). 
\remopt{This has clearly correlated with the sintered surface that changed depending on the layer (because the sample was printed horizontally).}
The duration of LTi's was comparatively long, and it fluctuated between \revnr{237,858} and \revnr{317,184} data points (\revnr{11.89} to \revnr{15.86~s}).
\remopt{We suspect that such significant fluctuation originates from the powder preheating to the set temperature close to the melting point, which is common practice with polymer PBF machines.}
As the slicing of the objects produced layers of equivalent size, we decided to use long periods of laser inactivity to distinguish between individual sintered layers. 
\remopt{This both simplified the code and improved its performance, as less data from only three actuators (laser and two galvanometers) had to be processed. } 

Even more critical was to understand whether the measured laser and galvanometer X and Y power side-channel data correlate with the points actually sintered on a specific layer.
In the sintering-relevant traces, we could clearly observe both the scanning strategy (i.e., the pattern in which the powder was sintered) used and the laser power levels used.
The side-channel data had a clear correlation with what we observed through the S2 camera installed inside the building chamber.

We will describe several other observations whenever we describe a special handling they required for our proposed reconstruction approach.

\subsection{General Concept of Object Reconstruction}

\begin{figure}[tbp!]
    \centering
    \includegraphics[width=.95\linewidth]{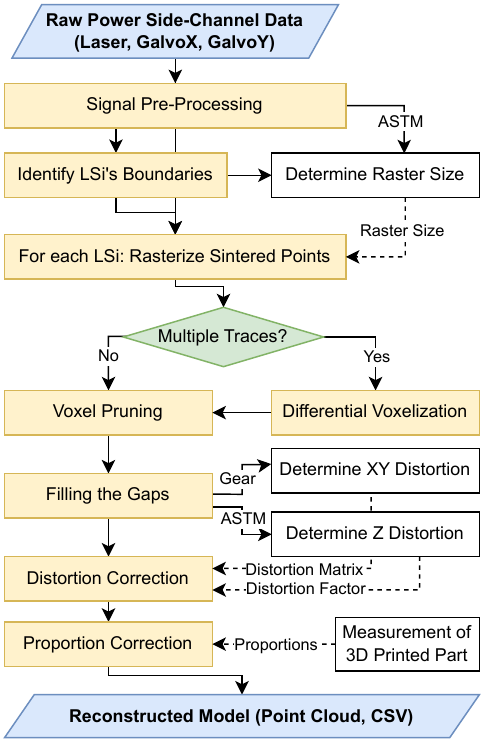}
    \caption{Workflow outlining individual steps of \emph{Power Side-Channel-based Reconstruction} of a model 3D-printed on PBF. \remopt{Unfilled (white) boxes indicate operations conducted only once to determine various reconstruction parameters. Filled (yellow) boxes are applied to all processed traces.} }
    \label{fig:general_concept}
\end{figure}

Figure~\ref{fig:general_concept} outlines the main stages of our approach for the reconstruction of 3D-printed models based on the information from the power side-channel.
First, the collected data undergo preprocessing, which was informed by our manual analysis of the raw traces and their characteristics.
Then, we identify layer sintering (LS) boundaries that correspond to individual layers of a produced part. 
We then used preprocessed traces of one ASTM specimen and information about its LSi's to determine the raster size.
\remopt{Then, this parameter is reused across all traces.}
For each preprocessed trace, we use information about its LSi boundaries and the determined Raster Size to rasterize the sintered points of each layer.
Combined, rasterization of all individual sintered layers describes 3D voxels of the printed part.
Each voxel has an associated ``hit count'' that represents the number of times this voxel was sintered by an active laser.

When we have multiple power side-channel traces collected for 3D prints of the same object (this was the case with the ASTM specimen), we can aggregate this information in what we call \emph{differential voxelization}. 
Whether from a single or combined from multiple traces, the voxelized representation undergoes voxel pruning, which removes potential ``false positive'' voxels.
Afterwards,we try to recognize the ``false negative'' voxel and fill the gaps.

The models reconstructed up until this point show distortions in the XY plane and along the Z-axis.
We use the voxelized representation of Gear to determine the XY distortion matrix, and of the ASTM specimen to determine the Z distortion factor.
Both of these parameters are used to correct the distortion in all reconstructed models.
Lastly, we can use the measurements of 3D-printed parts to further correct proportions of the reconstructed models.
In the final step, we can save the reconstructed 3D model as a point cloud.

\remopt{For interested readers, we illustrate major stages of this reconstruction process in Appendix~\ref{app:reconstruction_stages}.}

\subsection{Signals preprocessing}
\label{sec:reconstruction:filtering}

\begin{figure}[tbp]
    \centering
    \begin{lstlisting}[backgroundcolor = \color{lightgray!10}, frame = single, language = C] % , basicstyle=\tiny

NormalizeLaser (RawLaser, ThresholdON, ThresholdOFF)
{
  LaserState = OFF;
  
  forall (iDataPoint in RawLaser)
  {
      // Only transition state if threshold is broken
      // Otherwise, maintain previous state
      if (RawLaser[iDataPoint] >=ThresholdON) 
        LaserState = LASER_ON;
      else if (RawLaser[iDataPoint] <= ThresholdOFF)
        LaserState = LASER_OFF;
        
      // Save normalized state
      NormLaser[iDataPoint] = LaserState;
  }
    
  return NormLaser;
}

    \end{lstlisting}

    \caption{Normalizing laser power  to ON/OFF (1/0) state. Based on the preliminary manual signal analysis, the  ${LaserThresholdON}$ was set to to \revnr{2.2~V} and ${LaserThresholdOFF}$ was set to to \revnr{1.1~V}. Note that the threshold separation account for the noise that can occasionally ``break through'' the activation thresholds of a specific state.     }
    \label{fig:code:laser_normalize}
        
\end{figure}

During manual analysis of the collected raw power side-channel traces, we identified a certain amount of noise (which was expected).
Noise in the laser signal could impact the false positive and negative recognition of a sintered point.
Noise in the galvanometer signals could influence the identified ${(x,y)}$ position of the sintered point.
Our signal preprocessing aims to reduce such effects.
\remfirst{Furthermore, to simplify algorithms and improve the performance, we also normalize the laser signal from the measured power values to its ON/OFF (0/1) states.}

The laser power signal has a very clear and distinct characteristic. 
The ``ON'' state is centered around \revnr{2.5~V} and ``OFF'' is below \revnr{1~V}. 
However, in the raw traces we observed occasional noise spikes that could break through the thresholds.

We decided to use a simple algorithm with two thresholds to normalize the laser signal to ON/OFF values of 1 and 0 (see Figure~\ref{fig:code:laser_normalize}).
To account for most of the noise, we decided to slightly shift our thresholds to \revnr{2.2~V} (for the transition to the ON state) and to \revnr{1.1~V} (for the transition to the OFF state).
The ON and OFF thresholds are used to change the ${LaserState}$, otherwise the current state is used.
This allows us to compensate for occasional excessive noise that could break through the activation thresholds of the current state.
Should an occasional spike break through the opposite threshold, we expect it to be compensated by two means.
A spike that very briefly transitions to the ON state will be removed during the voxel pruning phase that ensures that possible ``false positives'' are removed. 
A spike that very briefly transitions to the OFF state will be compensated by rasterization, which assumes multiple sintering ``hits'' of a single raster, and later by the gap filling phase, which aims to reduce ``false negatives''.

We also decided to remove low-amplitude high-frequency noise from the GalvoX and GalvoY signals.
For this purpose, we applied to the raw signals a 4th order Butterworth low-pass filter (LPF) with the cutoff frequency at \revnr{6~kHz}.
Note that LPF might (and does) slightly shift the signal.
We decided to ignore this, as\remopt{, similar to the case of normalized laser signal,} it should be compensated by rasterization and voxel pruning.
\remopt{Even though we clearly oversampled the signal, we decided against downsampling because during the rasterization process it will mainly affect the ``hit count'' of the raster which in turn could allow more aggressive voxel pruning.}

\subsection{Identifying LSi's Boundaries}
\label{sec:reconstruction:LSi_boundaries}

\begin{figure}[tbp!]
    \centering

\begin{tikzpicture}[scale=0.15]
\tikzstyle{every node}+=[inner sep=0pt]
\draw [black] (24.2,-10.2) circle (3);
\draw (24.2,-10.2) node {$LT$};
\draw [black] (46.8,-23.2) circle (3);
\draw (46.8,-23.2) node {$LSi$};
\draw [black] (22.5,-30.2) circle (3);
\draw (22.5,-30.2) node {$LS2LT$};
\draw [black] (27.187,-10.463) arc (81.8793:38.30384:27.793);
\fill [black] (45.07,-20.75) -- (44.97,-19.81) -- (44.18,-20.43);
\draw (44.24,-13.38) node [above] {$Laser\mbox{ }=\mbox{ }ON,\mbox{ }i++$};
\draw [black] (14.9,-10.2) -- (21.2,-10.2);
\draw (14.4,-10.2) node [left] {$i\mbox{ }:=\mbox{ }0$};
\fill [black] (21.2,-10.2) -- (20.4,-9.7) -- (20.4,-10.7);
\draw [black] (46.143,-20.285) arc (220.42957:-67.57043:2.25);
\draw (52.02,-15.75) node [above] {$Laser\mbox{ }=\mbox{ }ON$};
\fill [black] (48.71,-20.91) -- (49.65,-20.77) -- (49,-20.01);
\draw [black] (44.795,-25.428) arc (-45.97382:-101.88641:21.545);
\fill [black] (25.38,-31.02) -- (26.06,-31.67) -- (26.27,-30.7);
\draw (43.23,-31.63) node [below] {$Laser\mbox{ }=\mbox{ }OFF,\mbox{ }Ctr\mbox{ }:=\mbox{ }0$};
\draw [black] (24.437,-27.913) arc (135.57119:76.56858:20.611);
\fill [black] (43.94,-22.29) -- (43.28,-21.62) -- (43.05,-22.59);
\draw (29.2,-21.75) node [above] {$Laser\mbox{ }=\mbox{ }ON$};
\draw [black] (23.823,-32.88) arc (54:-234:2.25);
\draw (22.5,-37.45) node [below] {$Laser\mbox{ }=\mbox{ }OFF,\mbox{ }Ctr++$};
\fill [black] (21.18,-32.88) -- (20.3,-33.23) -- (21.11,-33.82);
\draw [black] (20.173,-28.321) arc (-136.59358:-233.12335:11.204);
\fill [black] (21.59,-11.66) -- (20.65,-11.74) -- (21.25,-12.54);
\draw (16.53,-19.61) node [left] {$Ctr\mbox{ }>\mbox{ }1,000$};
\draw [black] (24.697,-7.253) arc (198.16235:-89.83765:2.25);
\draw (33.77,-4.23) node [above] {$Laser\mbox{ }=\mbox{ }OFF$};
\fill [black] (26.84,-8.8) -- (27.76,-9.03) -- (27.45,-8.08);
\end{tikzpicture}

	\caption{FSM to recognize boundaries of LSi's. Transition from LT to LSi indicates recognition of Layer Sintering start. Transition from LS2LT indicates recognition of Layer Sintering End. The counter ``i'' is the number of the layer.}
	\label{fig:FSM_LSi_recognition}
\end{figure}
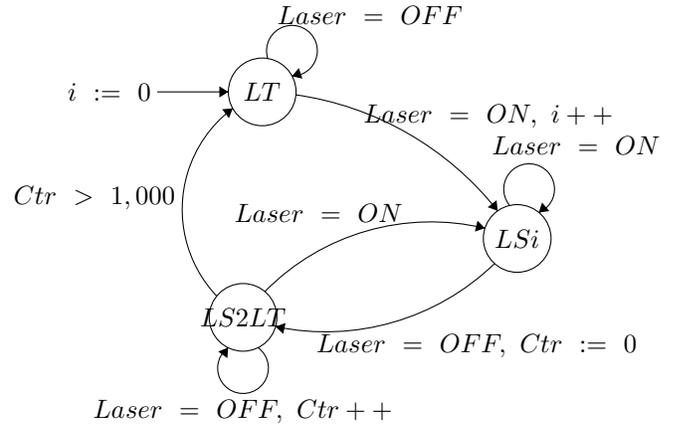

To identify LSi's boundaries, we implemented a simple finite-state machine (FSM) depicted in Figure~\ref{fig:FSM_LSi_recognition}.
This FSM consists of three states, Layer Transition (LT), Layer Sintering of i'th layer (LSi), and a state that represents the situation of uncertain transition from the LS to the LT state (LS2LT).
\remopt{The FSM starts in the LT state and initialize the layer counter ${i}$ to 0.
Scanning through the laser data, the FSM is in this state as long as the laser is OFF.}
When the laser turns ON, the FSM increments the layer counter and transitions to the LS state, where it stays as long as the laser is ON.
If the laser goes OFF, it does not necessarily mean the end of the layer sintering: This depends on the profile of the part and the scanning strategy employed by the machine.
Therefore, the FSM first transitions to an intermediate state LS2LT, in which it remains until one of the following two events occurs.
If the laser goes ON, it transitions back to the LSi state.
In the LS2LT state, each time it reads the laser OFF state, it increments the counter ${Ctr}$ (which was set to 0 during the transition to this state).
Only when the counter goes above \revnr{1,000}, the FSM recognizes this as the end of the layer sintering stage, and transition to LT.

Not shown in the FSM but implemented in our program, every time we transition to the LSi state, we mark the index of the data point (and thus know its timestamp) as the start of this layer.
Similarly, every time the FSM transitions from LS2LT to the LT state, we mark the last data point with laser ON (\revnr{1,000} data points before transitioning to LT) as the end of the layer. 
By the time we scan through all data points, the counter ${i}$ corresponds to the number of sintered layers.

\remopt{Note that this strategy and thresholding was sufficient for the PBF machine used and its specific timing. 
Working with other machines or using a different data acquisition rate might necessitate adjustments.
}

\subsection{Determining Raster Size}
\label{sec:reconstruction:raster_size}

Rasterization defines mapping the real coordinate values into discrete ``bins'' in the XY plane\remopt{~that can be addressed by their integer coordinates}.
Voxelization is a similar concept that describes ``binning'' in the 3D space, where the third dimension is along the build direction Z. 
Note that neither the voxel shape has to be cubic nor even the rasterization along the X and Y axes has to be identical. 
For a moment, we ignore the effects of selecting an unequal voxel height compared to the two other dimensions.
\remopt{We will revisit this in Section~\ref{sec:reconstruction:distortion_correction} where we deal with distortion correction.}

\begin{figure}[tbp]
  \centering
  \includegraphics[width=0.75\linewidth]{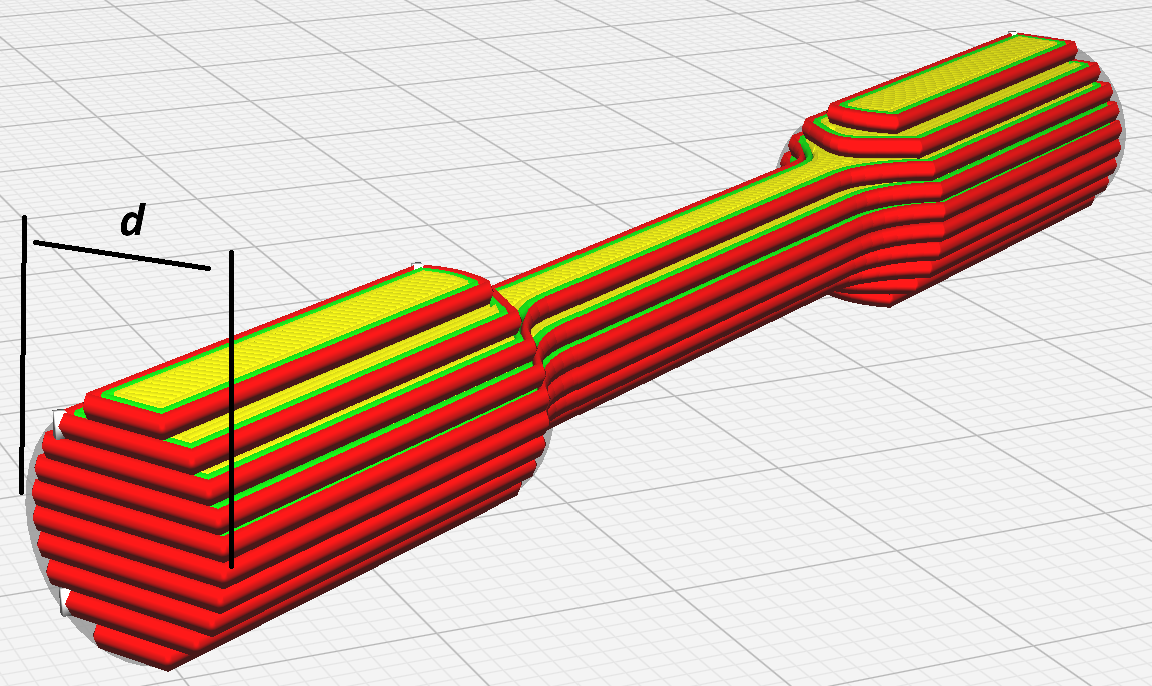}
  \caption{Sliced ASTM E8 (slicing is very coarse for the sake of visualization). Due to the cylindrical shape, the height and breadth are identical (equal to the cylinder diameter). 
  We exploit this fact for determining the raster size.
  }
  \label{fig:ASTM_E8-Sliced}
\end{figure}

We decided to have a quadratic rasterization, that is, the X- and Y-sides are of the same length. 
To determine the raster size, we took advantage of the fact that the ASTM E8 specimen was printed on the side. 
As it has a round base, we know that its maximum diameter ${d}$ in its middle layer should be equal to the height of the specimen that sums up across all printed layers (see Figure~\ref{fig:ASTM_E8-Sliced}).

The algorithm for determining LSi (described in Section~\ref{sec:reconstruction:LSi_boundaries}) provided us with the number of sintered layers and the boundaries in the traces where sintering begins and ends.
According to these data, the ASTM E8 print had \revnr{101} layers.
Consequently, layer \revnr{51} is the one with the supposed maximum diameter.

Based on our observations through the internal camera, the sintering pattern resembles a seesaw movement back-and-forth along one axis with comparatively slow advancement along another axis.
With this pattern, the laser had to traverse back-and-forth from one end of the specimen to another. 
In galvanometer signals, it is represented as a clear seesaw pattern that is more prominent on one than on another galvanometer.
We decided to exploit this fact to identify the difference between local maxima and minima in the galvanometer signal. 
The largest of such deviations should correspond to the distance traversed between two sintered points at opposite edges of the sample.

\begin{figure}[tbp]
    \centering
    \begin{lstlisting}[backgroundcolor = \color{lightgray!10}, frame = single, language = C] % , basicstyle=\tiny

DetermineMaxDeltaX (GalvoX_layer51)
{
  // Aggressive LPF to remove high-frequency noise remains 
  GalvoX_lpf = LowPassFilter(GalvoX_layer51, 20000, 1000);

  // Find peaks (local minima and maxima) in the signal
  Peaks_l51 = FindPeaks (GalvoX_lpf);

  // find max. distance between consequetive (min/max) peaks
  maxDeviation = 0;
  for (i=0; i<length(Peaks_l51)-1; i++)  
    if (maxDeviation < abs(Peaks_l51[i] - Peaks_l51[i+1]))
      maxDeviation = abs(Peaks_l51[i] - Peaks_l51[i+1]);
      
  return maxDeviation;
}

    \end{lstlisting}

    \caption{Determining maximum GalvoX deviation during the printing of a specific layer. The algorithm for GalvoY is similar. Both were applied to layer 51 of ASTM E8 specimen.}
    \label{fig:code:find_max_galvo_deviation}
        
\end{figure}

The pseudocode in Figure~\ref{fig:code:find_max_galvo_deviation} determines the maximum deviations of GalvoX during the seesaw pattern (the code for GalvoY is almost identical).  
\remopt{The only notable aspect of this fairly simple algorithm is that we first apply an aggressive low-pass filter at \revnr{1 kHz} cutoff that removes the remains of the high-frequency noise.
Subsequently, the local peaks represent the terminal positions of the seesaw pattern.}

After applying this algorithm to layer 51 of the ASTM E8 specimen, we received the following values. 
The maximum GalvoX deviation between the local maximum and minimum peaks was \revnr{0.1009~V}, and the corresponding value for GalvoY was \revnr{0.0228~V}\footnote{\remopt{We will discuss the meaning of measured voltage value in the next section.}}.
The specimen was printed not entirely along the X-axis, but at a slight angle, which complicated determining which galvanometer deviation would correspond to the diameter ${d}$.
Using the Pythagoras equation, we calculate the maximum ``length'' of movement corresponding to \revnr{0.1034~V}.
After dividing it by 101 (number of layers), we get the value of \revnr{0.001~V}, which should result in a cubical voxel (with equal length along all axes).
\remopt{We confirmed this assumption by processing ASTM E8 specimen with this raster size and receiving seemingly spherical cross-section when looking from the size.
However, we decided on a different raster size guided by the following considerations.}

Determining the size of the raster has several trade-offs.
Our primary motivation for the rasterization was to compensate for the remaining noise in the galvanometer signals \remopt{(we could not apply an \rev{aggressive} low-pass filter as it could distort the real signal).}
Considering the raster as ``hit'' if the laser sinters anywhere inside its boundaries introduces an average positional error of ${\pm 1/2}$  of the raster length.
Thus, a larger raster will \remopt{undoubtedly} increase the reconstruction error.
If the raster is too small, it could lead to gaps in the rasterized representation of the sintered layer.
This can be exacerbated by the occasional laser spikes that break through the opposite threshold values \remopt{(which we occasionally observed)}. 
With the larger raster, such spikes can be largely ignored, as every raster element should receive at least one laser hit.

\begin{figure}[tbp]
    \centering
    \begin{subfigure}[b]{0.5\linewidth}
        \centering
		\includegraphics[width=\linewidth]{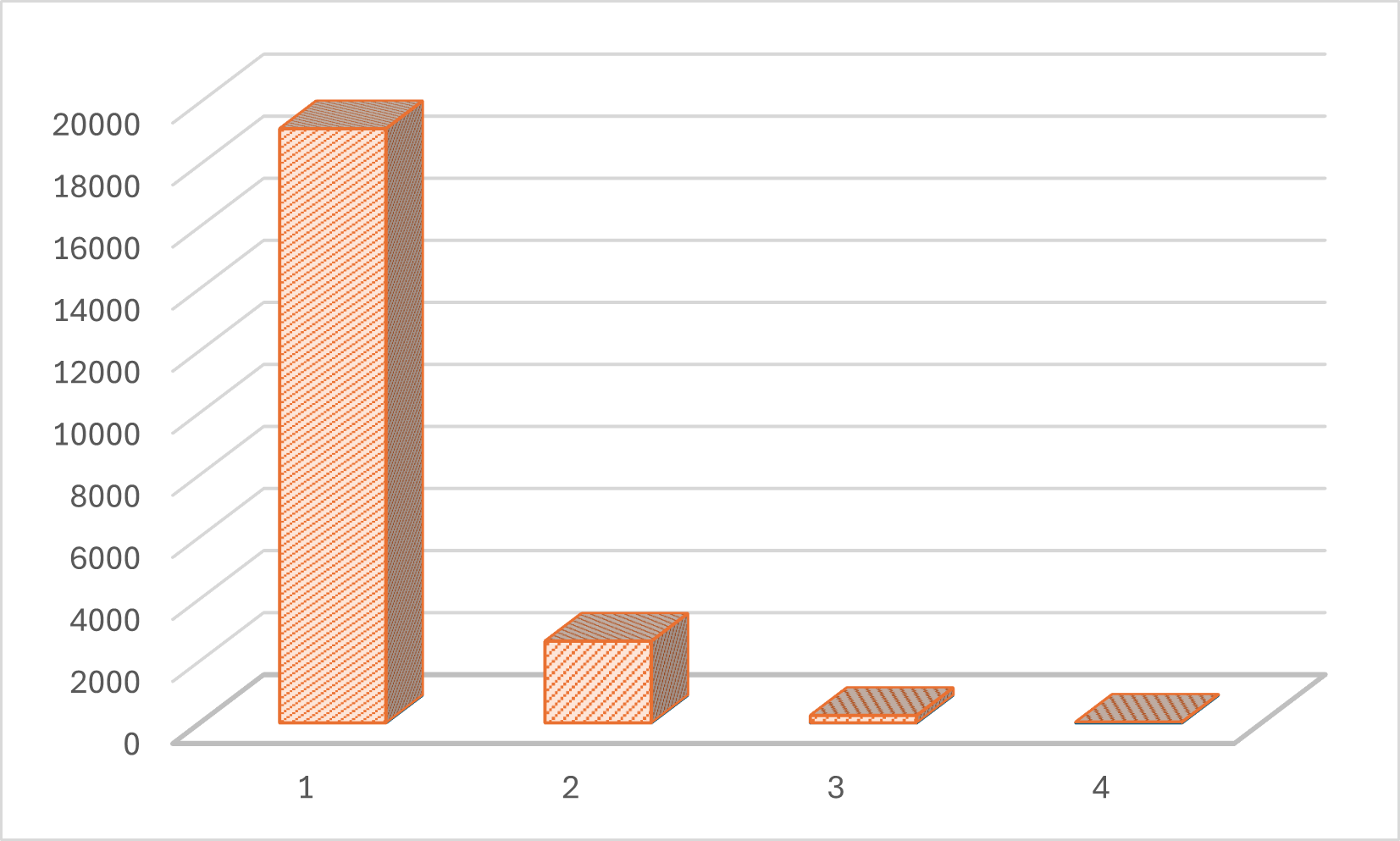}
        \caption{Raster size: 0.0010}
		\label{fig:ASTM2_layer51_HitCtrHist_Raster0010}
    \end{subfigure}%
    \begin{subfigure}[b]{0.5\linewidth}
        \centering
		\includegraphics[width=\linewidth]{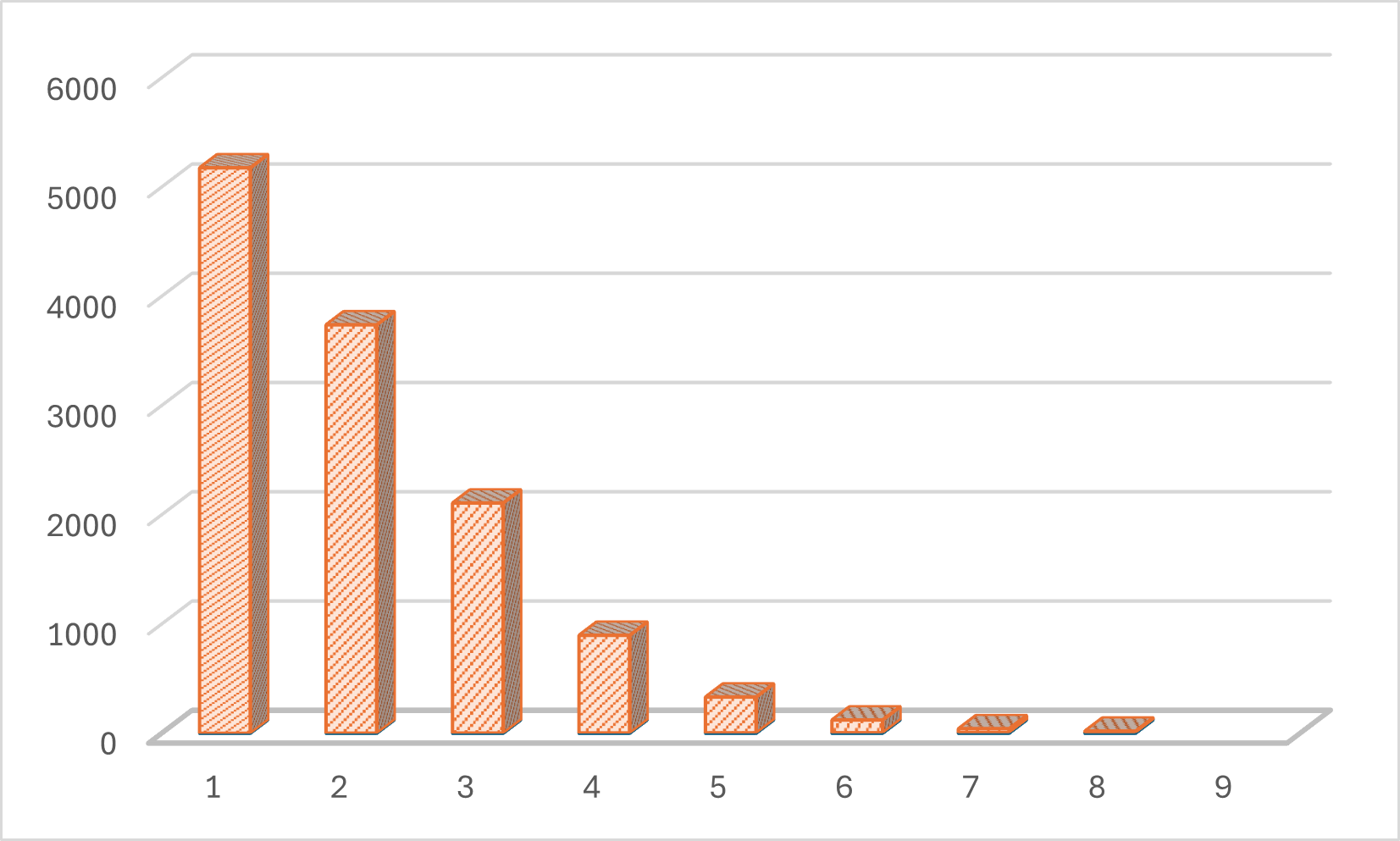}
        \caption{Raster size: 0.0025}
		\label{fig:ASTM2_layer51_HitCtrHist_Raster0025}
    \end{subfigure}%

	\caption{Histograms of the ${HitCtr}$ distribution for layer 51 of ASTM E8 specimen under selected raster size. 
    }
	\label{fig:ASTM2_layer51_HitCtrHist}
\end{figure}

To inform our decision, we rasterized ASTM E8 galvanometer data starting with \revnr{0.001~V} (as determined above, it should roughly correspond to ${1/101}$ of the diameter of the specimen) and gradually increasing the cell size with step size \revnr{0.0005~V}. 
Based on the histograms of the ``hit counter'' per raster (see two selected histograms in Figure~\ref{fig:ASTM2_layer51_HitCtrHist}) and the described trade-offs, we decided to use the length of the \revnr{raster cells of \revnr{0.0025~V}} for both GalvoX and GalvoY signals.

\remopt{Note that the discrepancy between raster length and layer height will cause distortion along the Z-axis when the ``hit'' (or sintered) voxels are represented in the coordinate space \emph{(RasterX, RasterY, LayerNr)}. 
The selected raster size will lead to the elongation along the Z-axis.
We will correct for this distortion in Section~\ref{sec:reconstruction:distortion_correction}.}

\subsection{Rasterization of Layer Sintered Points}
\label{sec:reconstruction:rasterization}

The initial challenge with the rasterization was that we did not know what the galvanometer measurements actually represent.
The two equally probable assumptions were that they correspond to (i) the actual coordinates or (ii) the angle at which the laser beam is deflected. 
According to the Sinttratec website, S2 has a round built plate of \revnr{160~mm} diameter.
We estimated that the distance from the built plate to the window through which the laser enters the build chamber (the galvanometers are situated slightly above) is at least \revnr{400~mm}. 

Given that the models were smaller than the build plate and positioned somewhat in the middle of it, we could confidently assume that the angle would never exceed ${\pm \pi/8}$.
In these ranges, the difference between the angle and its tangent is negligible (illustrated in the Appendix~\ref{app:diff_theta_tantheta}).
Therefore, we decided to treat the measurements as coordinates.
Note that we ignore height because it introduces a constant scaling factor between angle and position (${Pos = Height \cdot tan(\theta)}$), resulting only in the scaling of the reconstructed model.

Given that we interpret galvanometer readings as coordinates and ignore a possible scale factor, the rasterization of layer data is straightforward (see pseudocode in Figure~\ref{fig:code:rasterization}).
The function operates on the preprocessed data from laser and both galvanometers. 
It uses the identified LSi boundaries to iterate through all layers. 
For each sintered point (when the laser was ON), it uses ${RasterSize}$ (determined in Section~\ref{sec:reconstruction:raster_size}) to convert the galvanometer coordinates to the rasterized ones. 
Then, the ${HitCtr}$ of the specific voxel in the ${(RasterX, RasterY, LayerNr)}$ coordinate space is increased by one.
\remopt{The result is is the point cloud, where each point represents its voxel and ${HitCtr}$ indicates how many times this voxel has been hit by a laser.}

\begin{figure}[tbp]
    \centering

    \begin{lstlisting}[backgroundcolor = \color{lightgray!10}, frame = single, language = C] % , basicstyle=\tiny

RasterizeAllLayers (Laser, GalvoX, GalvoY, LSi)
{
  InitializeWithZero (VoxelLXYH);

  for (LayerNr=0; LayerNr<length(LSi); LayerNr++)
    for (i=LSi[Layer].start; i<=LSi[Layer].end; i++)
    {
      if (Laser[i] == LASER_ON)
      {
        // Rasterize X- and Y- coordinates
        RasterX = GalvoX[i] / RasterSize;
        RasterY = GalvoY[i] / RasterSize;

        // Increment "hit count" in a specific voxel
        VoxelLXYH[LayerNr][RasterX][RasterY]++;
      }
    }
}

    \end{lstlisting}
    
    \caption{Rasterization of Galvo X/Y measurements for all layers. 
    The result: Voxels in the coordinate space ${(RasterX, RasterY, LayerNr)}$ with an associated $HitCtr$.    }
    \label{fig:code:rasterization}
        
\end{figure}

\subsection{Differential Voxelization}
\label{sec:reconstruction:differential}

We take our inspiration for the proposed \emph{differential voxelization} from the work of Kocher et al.~\cite{kocher1999differential} \remopt{who introduced Differential Power Analysis (DPA) for breaking cryptographic keys}.
The fundamental idea is that multiple traces collected for the execution of the same process can be aggregated to amplify the analyzed signal.

We considered two options of how it can be applied to the object reconstruction.
First, the segments of raw (or preprocessed) traces corresponding to individual LSi's can be added up. 
As noise is generally random, this operation should amplify the real signal, thus increasing the Signal-to-Noise Ration (SNR) in the aggregated signal.
The prerequisite for this is that the traces are perfectly aligned, have exactly the same duration, and all operations happen at precisely the same time offsets.
To verify whether it is the case, we compared the durations of corresponding LSi's from two different traces of ASTM E8 specimen print. 
Although the duration of most layers deviated by just 1-2 poling intervals (which at 20 kHz poling frequency corresponded to 50-100~${\mu}s$) and the maximal deviation was at 6 poling intervals (300~${\mu}s$), we decided against this approach.

Another alternative, for which we decided, is to aggregate already rasterized (or voxelized) data points.
As after rasterization the real measurements of all traces are mapped into the same discrete set of coordinates (voxels), we can aggregate the laser ${HitCtr}$ numbers associated with them (see pseudocode in Figure~\ref{fig:code:diff_trace_aggregation}). 
This should amplify the ``hit counters'' of really sintered voxels while keeping them low for erroneous ones.
This could allow for the use of more aggressive parameters for voxel pruning, which we discuss next.

\begin{figure}[tbp]
    \centering
    \begin{lstlisting}[backgroundcolor = \color{lightgray!10}, frame = single, language = C] % , basicstyle=\tiny

DifferentialVoxelization (arrVoxelLXYH)
{
  InitializeWithZero (DiffVoxelLXYH); 
  
  // Iterate through all voxels in all voxelized traces
  forall (iTrace, iLayer, iPosX, iPosY) 
      DiffVoxelLXYH[iLayer][iPosX][iPosY] += 
        arrVoxelLXYH[iTrace][iLayer][iPosX][iPosY];

  return DiffVoxelLXYH;
}
    \end{lstlisting}

    \caption{Differential Trace: Aggregate ``Hit Counters'' of all traces collected for multiple prints of the same object. Note that the position, orientation, \remopt{and other printing parameters} should be the same. }
    \label{fig:code:diff_trace_aggregation}
        
\end{figure}

\subsection{Voxel Pruning}
\label{sec:reconstruction:voxel_pruning}

When we plotted the ``lit'' (${HitCtr > 0}$) voxels as point clouds,  we observed numerous spurious points. 
They appear like a ``halo'' of false positive voxels around the recognizable reconstructed model (illustrated in Appendix~\ref{app:halo}).
To improve the quality of reconstruction, these points had to be removed.

We implemented two strategies of voxel pruning.
First, voxels can be pruned according to their ${HitCtr}$. 
Second, voxels can also be deemed erroneous based on the absence of a sufficient number of neighbors in their proximity. 
The pseudocode describing the voxel pruning is shown in Figure~\ref{fig:code:voxel_prooning}.

To determine the parameters for both, we first collected several statistics on the rasterized representation of one of the ASTM E8 samples. 
In Section~\ref{sec:reconstruction:raster_size} we used statistic on ${HitCtr}$ (hits of laser inside the raster) to inform our decision on the raster size.
According to that statistic, under the selected raster of \revnr{0.0025~V} a substantial number of cells are hit only once (41.75~\% to be exact). 
For pruning based on \rev{differential voxelization} (as defined in Section~\ref{sec:reconstruction:differential}), we will revise this number to be at least \revnr{3} for the combination of three traces. 

For pruning based on the number of neighbors in the vicinity, we need first to specify what ``vicinity'' means and then determine two essential parameters: the distance from the voxel that defines the neighborhood and the minimal required number of neighbors. 
For performance reasons, we decided to use the vicinity in a 2D plane of an individual layer.
To define the parameters mentioned, we first calculated two different statistics on the rasterized representation of the ASTM E8. 

To determine the neighborhood distance, we calculated the length of the stretches of the ``gaps'', which we define as the number of unlit rasters ($HitCtr$ is 0) between those that are lit (${HitCtr >= 1}$).
We calculated these along the X-axis across all layers of the sample.
Based on the histogram of the length of these stretches (for the interested reader, we added them in Appendix~\ref{app:voxel_pruning}), we decided to have a neighborhood of \revnr{5}.
According to the collected statistics \revnr{97.1}~\% of gap stretches is shorter than \revnr{5}, increasing the probability that the ``true positive'' voxel will have enough neighbors at that distance.

\begin{figure}[tbp]
    \centering
    \begin{lstlisting}[backgroundcolor = \color{lightgray!10}, frame = single, language = C] % , basicstyle=\tiny

VoxelPruningByHitCtr (VoxelLXYH, ProoneThreshold)
{
  // Iterate through all voxels
  // Prune voxels whose "hit counter" is below the threshold
  forall (iLayer, iPosX, iPosY) 
    if (VoxelLXYH[iLayer][iPosX][iPosY] < ProoneThreshold)
      VoxelLXYH[iLayer][iPosX][iPosY] = 0;
}

VoxelPruningByNeighborNr (VoxelLXYH, Range, MinNeighborNr)
{  
  // Iterate through all voxels
  forall (iLayer, iPosX, iPosY) 
  {
    // skip voxels that have not been "hit"
    if (VoxelLXYH[iLayer][iPosX][iPosY] == 0)
      continue;

    // Compensate for "current" voxel
    Neighbours = -1; 

    // Count how many neighbour voxel "hit" in Range
    for (l=iLayer-Range; l<=iLayer+Range; l++)
      for (x=iPosX-Range; x<=iPosX+Range; x++)
        for (y=iPosY-Range; y<=iPosY+Range; y++)
          if (VoxelLXYH[l][x][y] > 0)
            Neighbours++;
            
    // Prune voxels than has insufficient close neighbors
    if (Neighbours < MinNeighborNr)
      VoxelLXYH[iLayer][iPosX][iPosY] = 0;
  }
}

    \end{lstlisting}

    \caption{Voxel Pruning. \remopt{Two strategies are used: based on minimal (1) voxel ${HitCtr}$, and (2) number of neighbors.} \todo{?? split in 2 sub-figures?}}
    \label{fig:code:voxel_prooning}
        
\end{figure}

We then used this distance to calculate the histogram of the number of neighbors across all lit voxels.
Based on these data, we decided to use \revnr{33} as the minimum number of neighbors, which would remove approximately \revnr{3.2~\%} of all voxels present in the analyzed sample.
Every voxel that has less than that will be removed from the reconstruction.
Note that this threshold is slightly higher than ${1/4}$ of the maximum possible neighbors at the distance (which is ${(2 \cdot 5+1)^2-1 = 120}$).
This might lead to false removal of the correct points at the edges. 
However, our experimentation showed that this is a fairly negligible trade-off: With these thresholds, the pruning algorithm removed the voxel halo, while not ``shaving off'' significant portions of the reconstructed model. 

For pruning of the differential reconstruction, we chose slightly more ``aggressive'' parameters.
Instead of distance \revnr{5}, we selected distance \revnr{4}, assuming that multiple combined traces would be more reliable to fill the gaps.
As the threshold for the minimal number of neighbors, we selected \revnr{22}, guided by similar considerations as described above.

\subsection{Filling the Gaps}
\label{sec:reconstruction:gap_filling}

Based on the gap analysis (conducted for the Voxel Pruning in Section~\ref{sec:reconstruction:voxel_pruning}), we identified that the reconstructed objects contained much more and much larger gaps than we anticipated.
As the models were printed with the full infill, we want to reflect this in the reconstructed model by ``filling the gaps''.
\remopt{However, the solution we developed and present in this section is not generalizable.
It depends on the geometry of the model itself and on the orientation in which it was printed.}

\begin{figure}[tbp]
    \centering
    \begin{lstlisting}[backgroundcolor = \color{lightgray!10}, frame = single, language = C] % , basicstyle=\tiny

Project_Gear (VoxelLXYH)
{  
  InitializeWithZero (ProjectedXYL, minHitCtr);
  
  // Iterate through all voxels
  forall (iPosX, iPosY) 
  {
    maxLayer = -1;
    sumHitCtr = 0;
    
    // find highest layer with ``lit'' voxel with same (x,y) 
    forall (iLayer)
      if (VoxelLXYH[iLayer][iPosX][iPosY] > 0 && 
          iLayer > maxLayer)
      {
        maxLayer = iLayer;
        sumHitCtr += VoxelLXYH[iLayer][iPosX][iPosY];
      }

    // Add the highest layer if projection found
    if (maxLayer > -1 && sumHitCtr >= minHitCtr)
      ProjectedXYL[iPosX][iPosY] = maxLayer;
  }

  return ProjectedXYL;
}

    \end{lstlisting}

    \caption{Simplified algorithm for projection of all points to the highest layer in the original point cloud. 
    Here, the result is a 2-dimensional array with values representing the maximum layer number of projection. \remopt{In a real implementation, we maintained three dimensions and assigned aggregate ${HitCtr}$ to the projected voxel.} \remopt{For ASTM specimen, the projection is both up and down, starting from the middle layer.} \todo{?? remove?} }
    \label{fig:code:project_gear}
        
\end{figure}

\remsecond{A simplistic strategy would be to go layer-by-layer and then, withing each layer, iterate row-by-row to identify the minimum and maximum positions of the lit voxels in the row.
Then all voxels between these two ``terminals'' could be automatically lit.
Depending on the orientation in which the model was printed, this approach would largely work for ASTM E8. 
However, in certain orientations it would connect points that should not be connected, such as between two larger diameter parts on the opposite side of the specimen.
With the Gear model, it is even worse: This approach would connect the space between some of the teeth and also fill the hole in the middle of the model.
And this is not even considering the fact that some false positive voxels can increase the negative effect of such filling.}

We came up with a\remsecond{n alternative} solution that takes advantage of the properties of individual models. 
We first observe that multiple layers should have sintered points with the same ${(x, y)}$ coordinates.
This means that if the coordinate is missing in one layer (e.g., due to an unfortunate timing of the polling), it might appear in other layers.
Should it be the case, we could fill the gap, while otherwise leaving it unfilled.
However, here we should take into account the geometry of the model and, in general, also the orientation with which the model was printed to decide \rev{which layers to ``consult''}.

\begin{figure}[tbp]
  \centering
  \includegraphics[width=0.75\linewidth]{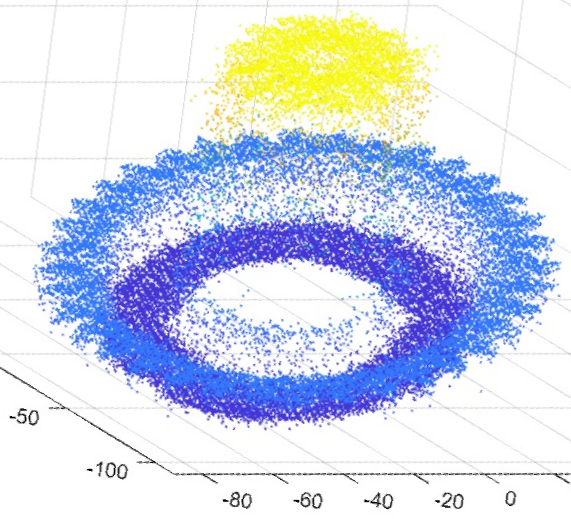}
  \caption{Reconstructed (voxelized) Gear model after each voxel was projected ``up'' to the single voxel with the same ${(x, y)}$ and largest $z$ coordinates. The ${HitCtr}$ is aggregated across all projected voxels with the same ${(x, y)}$. To avoid adding noise, voxels then pruned. \remopt{In the image, voxels pruned with the minimum ${HitCtr}$ threshold at 20.} \todo{replace with image of pruning by min. 55, as used in code!} }
  
  \label{fig:GEAR1_Rasterized_Projected}
\end{figure}

For the Gear, because there are no overhangs, we can ``project'' all voxels with the same ${(x, y)}$ coordinates on a single of those voxels with the maximal $z$ coordinate. 
We can then calculate the ${HitCtr}$ of this voxel as the sum of all the projected voxels.
As this is somewhat similar to the differential approach, we also aggressively prune the spurious voxels.
Note that the projection destroys the neighborhood relationships.
Therefore, the pruning can only be done on the basis of ${HitCtr}$. 
\remopt{A simplified pseudocode that only records the maximum ${LayerNr}$ at ${(x, y)}$ coordinates is shown in Figure~\ref{fig:code:project_gear}.}
The result of actually implemented projection is shown in Figure~\ref{fig:GEAR1_Rasterized_Projected}.
In our implementation, we applied the voxel pruning with a minimum ${HitCtr}$ of \revnr{20},  
preferring that some of the gaps remain unsealed in order to ensure that no additional noise (that is, false positive voxels) is added.

In the next step, to fill the gaps, we need to go through all unlit voxels in the boundaries of the reconstructed model.
If for the tested voxel there is a counterpart with the same coordinates ${(x, y)}$ in the previously generated projection and the projected layer is higher, then we can turn this voxel on by assigning it ${HitCtr=1}$.
Otherwise, we proceed to the next gap voxel.
As a measure to distinguish between the ``naturally lit'' and ``filled gaps'' voxels, in our implementation we also added 1 to the ${HitCtr}$ of all originally lit voxels.
The pseudocode for this approach is shown in Figure~\ref{fig:code:gap_filling}.

For the ASTM specimen, the filling of the gaps is slightly more complex.
Because of the natural overhangs caused by the round shape when printed on the side, simple projection up would yield wrong results.
To deal with this, we created two projections.
All layers below the middle layer are projected ``down'' and all layers above the middle layer are projected up, with the middle layer included in both of these projections (for interested readers, we added the view of this projection in Appendix~\ref{app:gap_filling}).
Also, the test had to be adjusted: for the bottom half of the layers, the test is whether the projected $z$ coordinate is smaller than that of the tested gap point, and for the upper layers, the condition for lighting the voxel is inverse.

\begin{figure}[tbp]
    \centering
    \begin{lstlisting}[backgroundcolor = \color{lightgray!10}, frame = single, language = C] % , basicstyle=\tiny

VoxelGapFilling_Gear (VoxelLXYH, ProjectedXYL)
{  
  // Iterate through all voxels
  forall (iLayer, iPosX, iPosY) 
  {
    // Increment HitCtr for all voxels already hit
    if (VoxelLXYH[iLayer][iPosX][iPosY] > 0)
      VoxelLXYH[iLayer][iPosX][iPosY]++;

    // Fill the gap if the layer is below the "projection"
    else if (iLayer < ProjectedXYL[iPosX][iPosY])
      VoxelLXYH[iLayer][iPosX][iPosY] = 1;
  }
}

    \end{lstlisting}

    \caption{Gap Filling for Gear model. The conditions under which voxel is filled are model-dependent. Both ``projection'' and algorithm are slightly different for the ASTM specimen.}
    \label{fig:code:gap_filling}
        
\end{figure}

\subsection{Distortion Correction -- XY \& Z}
\label{sec:reconstruction:distortion_correction}

Reconstruction of objects using the process described so far would yield figures that are clearly recognizable but also distorted (as exemplified in Figure~\ref{fig:GEAR_RasterizedPointCloud_XY-View}).
There are two distinct distortions.
The distortion in the XY plane probably originates from the slightly different electrical characteristics of the voltage-divider circuits that we built to instrument both galvanometers (see Section~\ref{sec:instrumentation}).
Another distortion, along the Z-axis, was the result of conscious choice of the raster size.
The way in which we approached the characterization and correction of both distortions is different.

\subsubsection{Correction of Distortion in XY Plane}

To determine the XY distortion, we exploited the fact that the Gear projection on this plane (i.e. looking at the model from a top) should be circular if its teeth could be disregarded. 
At the same time, the same projection of the reconstructed model resembled an elliptical shape.
Thus, we could reduce the problem as follows: Instead of solving this for the original (perfect) and reconstructed (distorted) models, we would be doing this for a circle and a tightly fitting ellipsoid.

\begin{figure}[tbp]
    \centering
    \begin{subfigure}[b]{0.45\linewidth}
        \centering
		\includegraphics[width=\linewidth]{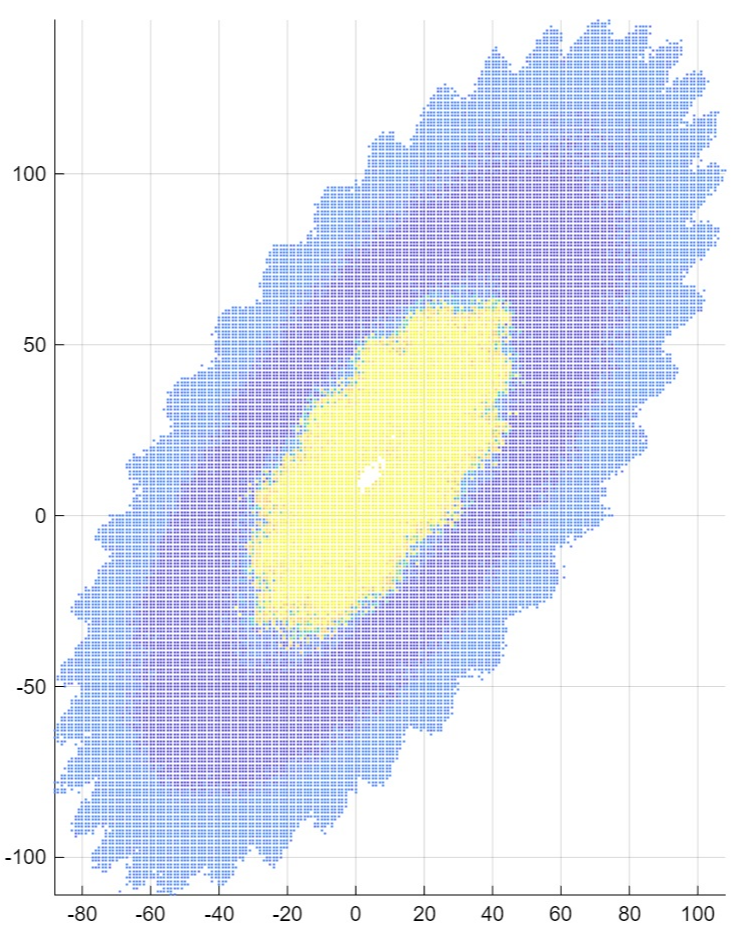}
        \caption{Distortion in XY Plane}
		\label{fig:GEAR_RasterizedPointCloud_XY-View}
    \end{subfigure}%
~
    \begin{subfigure}[b]{0.45\linewidth}
        \centering
		\includegraphics[width=\linewidth]{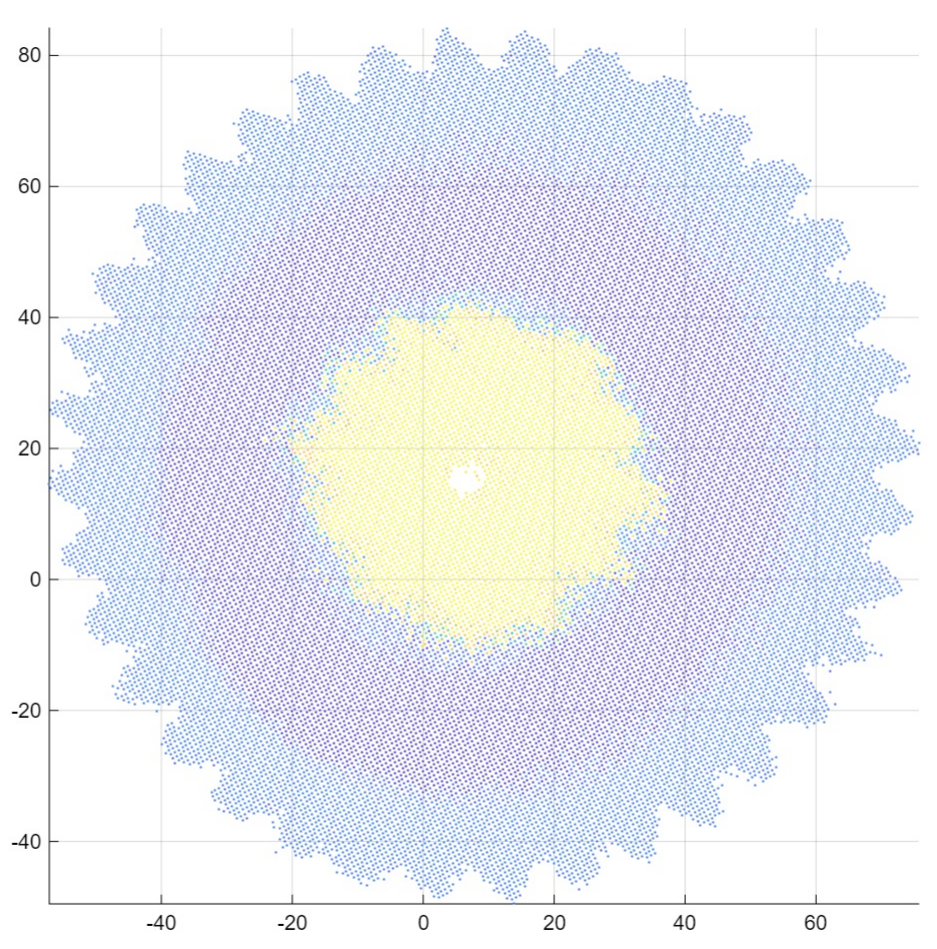}
        \caption{After distortion correction}
		\label{fig:GEAR_RasterizedCollapsedPruned_PointCloud_DistortionCorrected_XY-View}
    \end{subfigure}%

	\caption{XY view at the voxelized point cloud of Gear that we prepared for identification of the XY distortion factor\remopt{~(color indicates the height, that is the Z value)}.}
	\label{fig:GEAR_XY_Distorted_Corrected}
\end{figure}

As we used a single trace to reconstruct the Gear model, spurious (false positive) points could influence the definition of the fitting ellipse.
To deal with this situation, we used the projected representation of the model, as defined and used to fill the gaps in Section~\ref{sec:reconstruction:gap_filling}.
The resulting point cloud projected on the XY plane is shown in Figure~\ref{fig:GEAR_RasterizedPointCloud_XY-View}.



We disregard the $z$ coordinate of the projection, considering that all ${(x, y)}$ points are on a single plane.
There are established solutions to enclose such a point cloud in an ellipsoid~\cite{fitzgibbon1996direct, todd2016minimum}. 
In short, for a given set of points in 2D coordinates, the minimal fitting ellipse can be approximated by performing Principal Component Analysis (PCA) on the centered data. 
The approach determines the orientation and scale of the ellipse. 
When applied to our set of points, the algorithm yielded the following results.
The center of the point set was located at \revnr{(9.0864, 17.4401)},
the length of the major and minor axes of the fitted ellipse were \revnr{156.7775} and \revnr{67.5955}, respectively,
and the orientation angle was \revnr{0.9765}.



Determining and correcting for linear distortion from a reference circle to an ellipse in a rotated frame is another problem with a well-known solution~\cite{hartley2003multiple}.
\remfirst{It is widely used in applications like camera calibration or correction of optical distortions.}
As the scale factor of the reconstructed object is of no relevance (the original design in STL does not provide real dimensions anyway), we simple used a circle of radius \revnr{100} in the rasterized coordinates. 
We then determined the distortion parameters using this reference circle and the fitting ellipse \remopt{(an interested reader can find an illustration in Appendix~\ref{app:determining_distortion})}.
We calculated the following distortion:

\begin{math}
DistortMatrix =
    \begin{bmatrix}
        0.9556 & 0.4137 \\
        0.4137 & 1.2881
    \end{bmatrix}
\end{math}


If applied to the fitting ellipse, the inverse distortion matrix ${DistortMatrix^{-1}}$ will result in the reference circle.
To correct for the same distortion of an arbitrary point, the inverse affine transformation is applied after subtracting the center of the ellipse $C$ and then adding it back after the inverse: 

${px_{x,y}^{corr} = DistortMatrix^{-1}(px_{x,y}^{dist}-C)+C}$

We applied this approach to all reconstructed models, that is, to one Gear and three ASTM specimens reconstructed using the simple approach (i.e., single trace) and one ASTM specimen reconstructed using the differential approach.
The result of application to the Gear model is shown in Figure~\ref{fig:GEAR_RasterizedCollapsedPruned_PointCloud_DistortionCorrected_XY-View}.

\remopt{Note that this method restores the geometry of the distorted object under the assumption of global linear distortion.
Based on the results achieved by applying the same inverse transformation to all models processed to this stage, we consider this to be an appropriate approach.}

\subsubsection{Correction of Distortion along Z-Axis}

To identify and correct for distortion along the Z-axis, we used the ``differential'' reconstitution of the ASTM specimen\remsecond{, which underwent all the steps described above (that is, voxel pruning, gap filling, and XY distortion correction)}.
First, we observe that ASTM is elongated along one direction.
We used the coordinates ${(x, y)}$ of every point to compute a linear approximation (to a straight line that goes along the specimen) using a least squares polynomial fit of degree 1~\cite{seber2003linear}.
The angle of inclination ${\theta}$ of this line relative to the X-axis can then be calculated as ${\arctan}$ of the slope. 
The result was \revnr{${\theta = 0.4165}$} radian.

We used this angle to define the rotation matrix ${R}$.
By multiplying this matrix by the coordinates ${(x, y)}$ of every point in the reconstructed point cloud, we rotate the ASTM specimen parallel to the X-axis. 

Afterwards, we recall the fact that the ASTM specimen is supposed to have a round cross section.
This means that the widths and heights of the cross section should be identical.
After rotating this specimen parallel to the X-axis, the ratio is determined by the maximum deviations along the Y and Z axes.
The resulting \revnr{${deltaY = 53.5345}$} and \revnr{${deltaZ = 101}$} (in raster coordinates) mean that the elongation factor along the Z-axis is \revnr{1.8866}. 
To correct for this, we only need to divide the $z$ coordinates of all points by this factor.

\begin{table}[]
    \centering
    \begin{tabular}{|l|c|c|}
         \hline
        \textsc{Sec. \& Parameter} &  \textsc{Simple} & \textsc{Diff.} \\
         \hline
         \hline
         
         \textbf{\ref{sec:reconstruction:filtering}: Laser ThresholdON} & \multicolumn{2}{c|}{2.2V} \\ 
         \hline
         
         \textbf{\ref{sec:reconstruction:filtering}: Laser ThresholdOFF} & \multicolumn{2}{c|}{1.1V} \\ 
         \hline
         
         \textbf{\ref{sec:reconstruction:filtering}: Galvo -- LPF Cutoff 
         } & \multicolumn{2}{c|}{6 kHz} \\          
         \hline
         
         \textbf{\ref{sec:reconstruction:raster_size}: Galvo X/Y Raster Size} & \multicolumn{2}{c|}{0.0025V} \\ 
         \hline
         
         \textbf{\ref{sec:reconstruction:voxel_pruning}: Pruning -- Min Hit Ctr } & 1 & 3 \\ 
         \hline
         
         \textbf{\ref{sec:reconstruction:voxel_pruning}: Pruning -- Range } & 5 & 4 \\ 
         \hline
         
         \textbf{\ref{sec:reconstruction:voxel_pruning}: Pruning -- Min Neighbors} & 33 & 22 \\ 
         \hline

         \textbf{\ref{sec:reconstruction:distortion_correction}: XY Inverse Distortion Matrix } & \multicolumn{2}{c|}{
         \begin{math}
            \begin{bmatrix}
                1.2155 &  -0.3904 \\
               -0.3904 &   0.9017
            \end{bmatrix}
        \end{math}          
         } \\ 
         \hline

         \textbf{\ref{sec:reconstruction:distortion_correction}: Z Distortion Factor } & \multicolumn{2}{c|}{\revnr{1.8866}} \\ 
         \hline

         \textbf{\ref{sec:reconstruction:proportion_correction}: Proportion Correction - ASTM E8: ${l/d}$} & \multicolumn{2}{c|}{\revnr{6.7222}} \\ 
         \hline
         
         \textbf{\ref{sec:reconstruction:proportion_correction}: Proportion Correction - Gear: ${d/h}$} & \multicolumn{2}{c|}{\revnr{3.8629}} \\ 
         \hline
         
         \hline
         
    \end{tabular}
    \caption{Parameters used in reconstruction. }
    \label{tab:reconstruction_parameters}
\end{table}

\subsection{Proportion Correction}
\label{sec:reconstruction:proportion_correction}

When we reconstructed models using all the steps described above, we noticed that the reconstruction did not maintain the proportions of the original models. 
As malicious manufacturers will also have access to the physical 3D-printed parts, even simple measurements with a caliper can be used to determine its proportions.
Therefore, the reconstructed model can be corrected to comply with these proportions.

\rev{As we had access to the STL designs,} we used these to determine the proportions.
For the ASTM E8 model, we used the relationship between the length and the maximum diameter of the round part, which resulted in \revnr{6.7222}.
For the Gerar model, it is the relationship between the diameter of its base and the height, which resulted in \revnr{3.8629}. 
After determining the proportions, we scaled the reconstructed models (unevenly along the axes) to match these. 
We then saved the results as point clouds in CSV files for visualization and evaluation of the quality of reconstruction.

\section{Evaluation of Reconstruction}
\label{sec:evaluation}

\begin{figure}[tbp]
    \centering
    \begin{subfigure}[b]{0.49\linewidth}
        \centering
		\includegraphics[width=\linewidth]{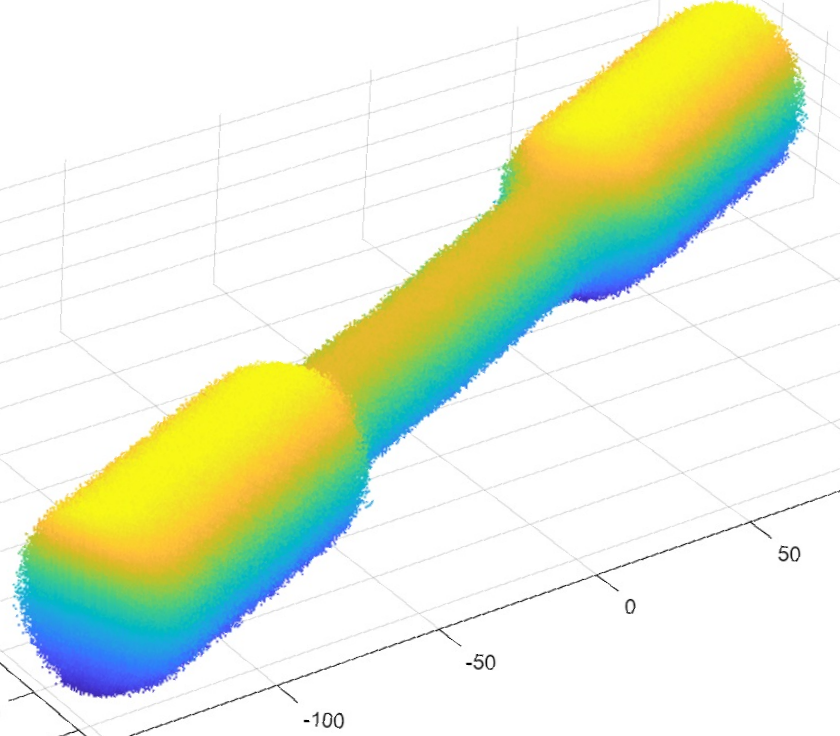}
        \caption{ASTM E8}
		\label{fig:recovered_model_ASTM}
    \end{subfigure}%
~~~
    \begin{subfigure}[b]{0.49\linewidth}
        \centering
		\includegraphics[width=\linewidth]{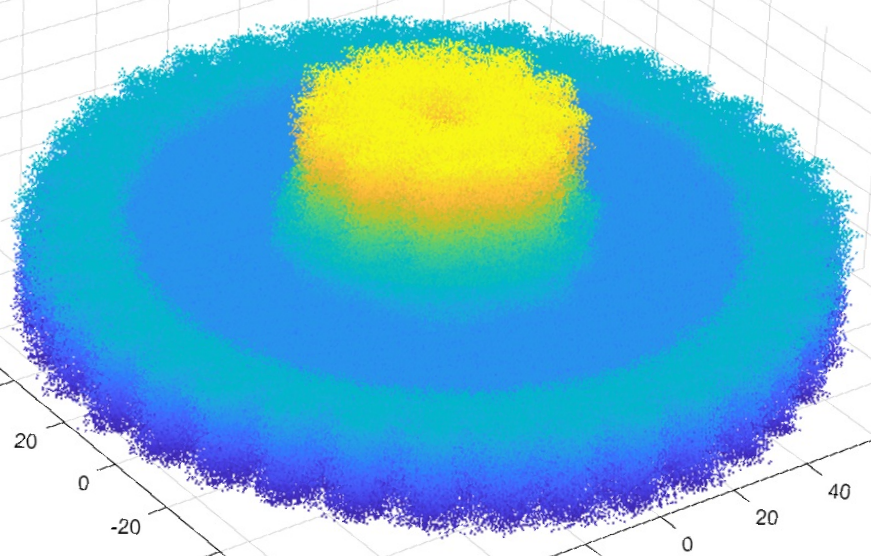}
        \caption{Gear}
		\label{fig:recovered_model_gear}
    \end{subfigure}%

	\caption{Reconstructed models (as Point Clouds). \remfirst{-- after going through all stages: (1) Signal preprocessing, (2) Rasterization, (3) Voxel Pruning, (4) Gap Filling, (5) Distortion Correction XY \& Z.} \todo{replace with views of models after proportion correction}  }
	\label{fig:reconstruction_results}
\end{figure}

We implemented the entire concept described in Section~\ref{sec:reconstruction} \remsecond{as well as the evaluation of reconstruction} in Matlab \remsecond{(Images of major reconstruction stages can be found in the Appendix~\ref{app:reconstruction_stages})}. 
Table~\ref{tab:reconstruction_parameters} summarizes the main parameters used in the reconstruction. 
Qualitatively, the reconstructed models appear to be very close to the original ones (see Figure~\ref{fig:reconstruction_results}).

To quantitatively assess the quality of reconstruction, we decided to evaluate volumetric True Positives, False Positives, and False Negatives. 
We voxelized the STL model using the equidistant grid size \revnr{0.25}\remopt{, which produced clearly recognizable representation}. 
We then re-positioned and scaled down the reconstructed models to fit closely. 
\remfirst{For Gear, we used the diameter of the base for calculating the scale factor.
For ASTM E8, we averaged the scale factors across all three dimensions.}
Afterwards, we re-voxelized the reconstructed model on the same grid used for STL models.
Lastly, we applied the algorithm shown in Figure~\ref{fig:code:assess_quality} to assess the quality of reconstruction. 
The results are summarized in Table~\ref{tab:reconstruction_results}.
For the interested reader, we add additional images that illustrate the True Positive, False Negative, and True Negative results in the Appendix~\ref{app:comparison}.

\remfirst{We analyze the factors that influenced the reconstruction results.
The Gear model turned out to be difficult to reconstruct.
Two factors contributed to rather bad False Positives.
First, the teeth in the reconstructed and STL models were slightly offset, probably because of a slight rotation during the printing.
Second, the applied Z distortion correction has not worked well for this model, as we observed that the reconstructed model was slightly higher than the STL one. 
With the ASTM specimen, the large False Negative results originate from the fact that, under the scale factor used, the reconstructed specimen was substantially shorter than the STL version.
}

We should note that, compared to the reconstraction based on a single trace, the proposed \emph{Differential} approach has led to a significant improvement in all the quality metrics evaluated.

\begin{figure}[tbp]
    \centering
    \begin{lstlisting}[backgroundcolor = \color{lightgray!10}, frame = single, language = C] % , basicstyle=\tiny

AssessQualityOfReconsruction (voxelizedSTL, voxelizedREC)
{  
  InitializeWithZero (TruePos, FalsePos, FalseNeg);

  // generate Point Clouds for True/False Positive/Negative
  forall (iLayer, iPosX, iPosY) 
  {
    if (voxelizedSTL[iLayer][iPosX][iPosY] > 0 && 
        voxelizedREC[iLayer][iPosX][iPosY] > 0) 
      TruePos[iLayer][iPosX][iPosY] = 1;
      
    else if (voxelizedSTL[iLayer][iPosX][iPosY] > 0 || 
             voxelizedREC[iLayer][iPosX][iPosY] > 0) 
      if (voxelizedSTL[iLayer][iPosX][iPosY] > 0)
        FalseNeg[iLayer][iPosX][iPosY] = 1;
      else
        FalsePos[iLayer][iPosX][iPosY] = 1;
  }

  // Count Voxels
  nrTruePos  = Count(TruePos);
  nrFalsePos = Count(FalsePos);
  nrFalseNeg = Count(FalseNeg);
  
  // Assess percentage
  percentTruePos  = nrTruePos  / count(voxelizedSTL) * 100; 
  percentFalsePos = nrFalsePos / count(voxelizedSTL) * 100; 
  percentFalseNeg = nrFalseNeg / count(voxelizedSTL) * 100; 
}

    \end{lstlisting}

    \caption{Generate point clouds of and assess True Positive, False Positive, and False Negative voxels and percentage. } 
    \label{fig:code:assess_quality}
        
\end{figure}


\section{Discussion \& Future Work}
\label{sec:discussion}

Several lessons and open research questions can be derived from the current work.

\paragraph{Implication on AM Outsourcing}
our work is the first that applies to industry. It proves that even significantly more complex AM machines used in industrial production are not immune to side-channel attacks.
The implications for the customers of AM outsourcing models are crystal clear: An ``analogue hole'' present in AM makes the security of outsourced design technically unattainable as is (and mitigation and enhanced trust in the printing vicinity are needed). 

\paragraph{Implication for OEMs}
In metal AM, the combination of process parameters (which are both numerous and material-dependent) is the close-guarded secret of the machine's Original Equipment Manufacturer (OEM). 
Our work shows that the power side-channel (as measured in our work) provides sufficient information for reverse engineering of parameters like the scanning strategy and laser power levels, which are among the most influential in this ``secret sauce''.

\paragraph{Sabotage Detection, Localization, and Investigation}
Another significant security threat in AM is \emph{sabotage}~\cite{yampolskiy2018security}, including part sabotage that can lead to the destruction of the system on which it is installed~\cite{belikovetsky2017dr0wned}.
Attack methods include the manipulation of the geometry of the part~\cite{sturm2014cyber, belikovetsky2017dr0wned, zeltmann2016manufacturing, graves2021sabotaging, moore2017implications} or of the process parameters used in the AM process~\cite{yampolskiy2015security, zinner2022spooky}.
 
Several researchers have proposed to detect such attacks based on side-channel measurements and demonstrated the feasibility of this approach on FDM desktop 3D printers~\cite{albakri2015non, chhetri2016kcad, belikovetsky2019digital, bayens2017see, gatlin2019detecting}.
The quality of reconstruction achieved in our work implies that it should also be possible for industrial-grade PBF machines.
Depending on the instrumented side-channels, it should be possible to detect both part geometry manipulations and at least some process parameter manipulations.
We are especially interested in combining sabotage detection with the presented reconstruction, which would allow \emph{side-channel based investigation} of sabotage\remopt{~by highlighting deviations}. 
Similarly to what Tsoutsos et al.~\cite{tsoutsos2017secure} demonstrated using the digitally obtained toolpath, we believe that side-channel-based reconstruction can also be used to assess the impact \remopt{of introduced manipulations} on parts performance.

\begin{table}[]
    \centering
    \begin{tabular}{|l|c|c|c|}
         \hline
        \textsc{Model} &  \textsc{True Pos.} & \textsc{False Pos.} & \textsc{False Neg.} \\
         \hline
         \hline


         \textbf{Gear} & 267266 & 108901 & 28752 \\ 
                       & (90.29~\%) & (36.79~\%) & (9.71~\%) \\
         \hline

         
         \textbf{ASTM run 1} & 210133 & 24973 & 54913 \\ 
          & (79.28~\%)  & (9.42~\%) & (20.72~\%) \\ 
         \hline
         
         \textbf{ASTM run 2}  & 200027 & 21594 & 65019 \\ 
          & (75.47~\%)  & (8.15~\%) & (24.53~\%) \\ 
         \hline
         
         \textbf{ASTM run 3} & 205929 & 20871 & 59117 \\ 
          & (77.7~\%)  & (7.87~\%) & (22.3~\%) \\ 
         \hline
         
         \textbf{ASTM Diff.}  & 214289 & 18602 & 50757 \\ 
          & (80.85~\%)  & (7.02~\%) & (19.15~\%) \\ 

         \hline
         \hline

         \textbf{Best Results} & 90.29~\% & 7.02~\% & 9.71~\% \\ 
         
         \hline
         
         \hline
         
    \end{tabular}
    \caption{Reconstruction Results. 
    True Positives are voxels present in both original and reconstructed model.
    False Positives are voxels present in the reconstructed but not present in the original model.
    False Negatives are, vice versa, voxels present in the original but not present in the reconstructed model.
    We give all results in the number of voxels as well as in the percentage in relation to the voxels in the original model. 
    Number of voxels in Gear model: \revnr{296018}, in ASTM E8 model: \revnr{265046}.
    Note that the proposed \emph{Differential} approach improved all reconstruction quality metrics assessed.  
    }
    \label{tab:reconstruction_results}
\end{table}

\remfirst{
\paragraph{Different AM (and non-AM) Processes}
\remopt{Due to various trade-offs, different AM technologies find different preferred areas of application.
The PBF technology investigated in the article dominates the manufacturing of ``net-shaped'' parts, but is limited in the volume of parts and the manufacturing speed.  
Direct Energy Deposition (DED) has no part size limitation, but requires substantial post-processing.
Another increasingly used technology is Cold Spray (CS) which can be used for part repair. 
There are numerous other computerized non-AM manufacturing technologies, such as multi-axis computer numerical control (CNC) machines.
Despite their role in the manufacturing industry, they have been largely neglected by security research.
We think that side-channels can provide an effective tool for both attacking and defending these machines.}
}


\section{Conclusion}
\label{sec:conclusion}

In this work, we have proven that it is possible to reconstruct designs 3D-printed on industrial-grade PBF machines using only a power side channel. 
To the best of our knowledge, to date this type of attack has only been shown against very simple FDM 3D printers.
Thus, our work has a range of profound consequences, both concerning and promising.

To start with, our attack means that even very complex industrial-grade machines such as PBF have an ``analogue hole'' that can be effectively exploited to bypass any employed cyber-security measures.
This is a significant finding because it should be taken into account in the risk assessment: external risk for design outsourcers, and internal risk for printer owners.
Furthermore, the side-channel attack can also be used to reverse engineer the manufacturing process parameters\remopt{~(such as scanning strategy, to name just one)}, which in metal AM is often considered a close-guarded ``secret sauce'' of OEMs.

On the bright side, the same side channels (when recorded privately) can provide a powerful basis to monitor and defend against another severe security threat in AM, sabotage.
It seems that it should be possible not only to detect sabotage but also to localize it in the part, and maybe even assess its impact on the part quality.

\section*{Code and Data Release}

Since this paper describes a very powerful attack, we will \emph{not} publicly release the code developed.
\remsecond{Researchers are welcome to contact the authors with the request for the code, although we reserve the right to decline such requests.}

To ensure independent verification, we release the data (from raw signals, through major intermediate stages, to the final verification results).
The data can be found under~\cite{anonymous_author_s_2025_16753249}. \todo{upload files and add link}


\section*{Acknowledgement}

This work was funded in part by the U.S. Department of Commerce, National Institute of Standards and Technology (NIST) under Grant NIST-70NANB21H121.

\clearpage

%% file: TDTonPBF_8_Appendix.tex
\appendices

\section{Difference between ${\theta}$ and ${tan(\theta)}$}
\label{app:diff_theta_tantheta}

\begin{figure}[h!]
  \centering
  \includegraphics[width=0.75\linewidth]{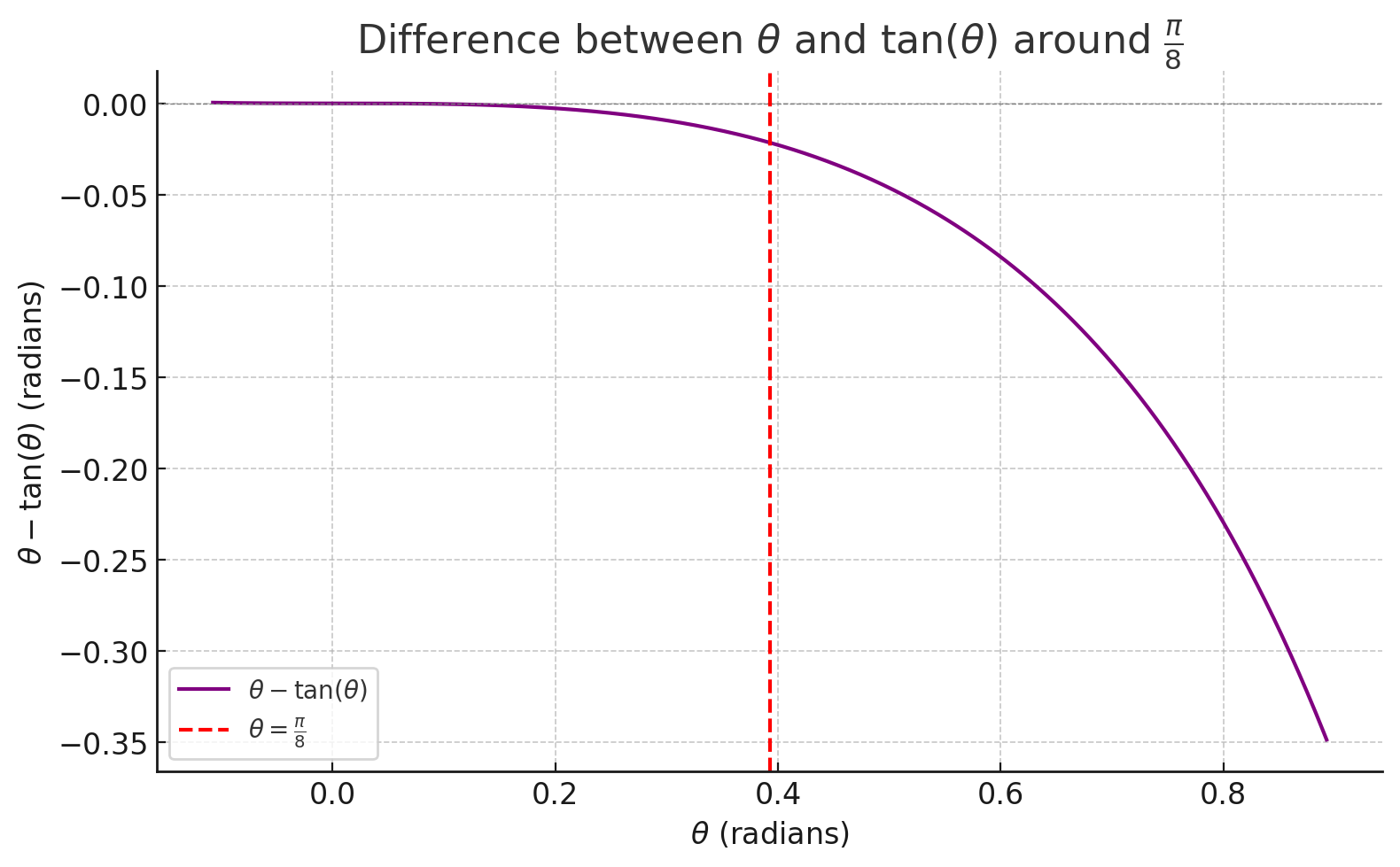}
  \caption{Difference between ${\theta}$ and ${tan(\theta)}$. Given that the diameter of the S2 build plate is 160 mm and the distance between laser source and the print bed at lease \revnr{400 mm}, the maximum angle should be well within ${\pm \pi/8}$. Therefore, we can use GalvoX and GalvoY measurements as substitute for actual coordinates, neglecting the possible scale factor of the height. }
  \label{fig:TanTheta_radian_Pi8}
\end{figure}

In Section~\ref{sec:reconstruction:rasterization} we had to decide how to interpret galvanometer signals, as an angle or as a position.
Figure~\ref{fig:TanTheta_radian_Pi8} illustrates that, at small angles, the difference is negligible.  

\section{Matters of Voxel Pruning}
\label{app:halo}
\label{app:voxel_pruning}

\begin{figure}
  \centering
  \includegraphics[width=.5\linewidth]{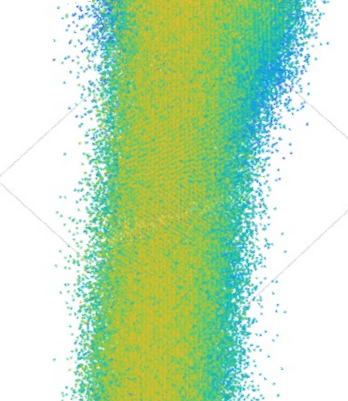}
  \caption{Using raw side-channel signals directly for reconstruction of printed object can lead to erroneous points.  
  Signal preprocessing (guarding laser signal against occasional spikes and removing high-frequency noise from galvanometer signals) and voxel pruning are working in unison to remove such unwanted points.  }
  
  \label{fig:ASTM2_Rasterized_zoom-in_halo}
\end{figure}

\begin{figure}
  \centering
    \includegraphics[width=.95\linewidth]{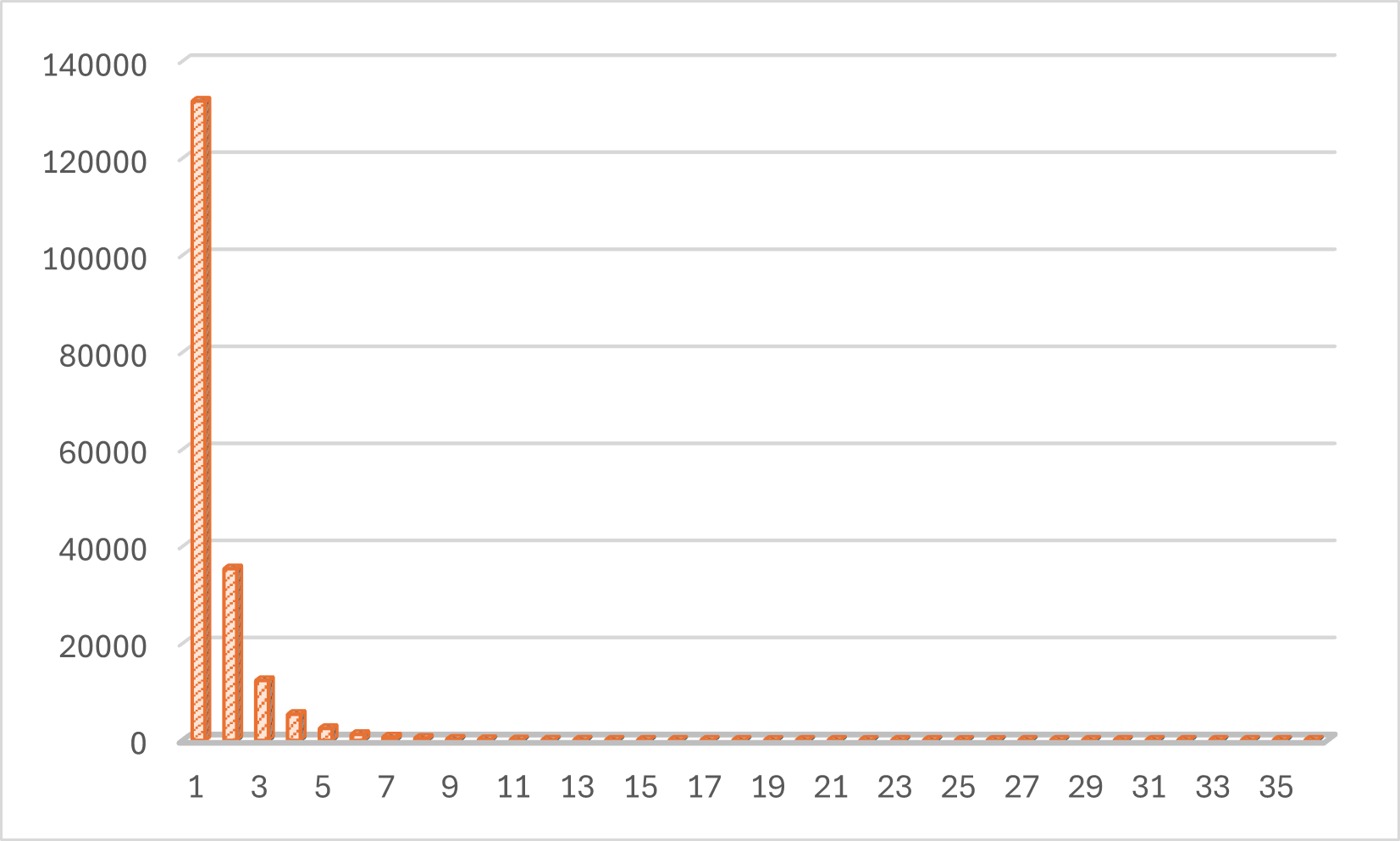}
    \caption{Length of Gaps stretches (number of not ``hit'' voxels) counted along the X-axis. The longest gap stretch is 36 voxels. The length of \revnr{97.1~\%} of all stretches is shorter than \revnr{5}. }
    \label{fig:ASTM_Stat_GapStretchHist}
\end{figure}

\begin{figure}[ht!]
  \centering
    \includegraphics[width=.95\linewidth]{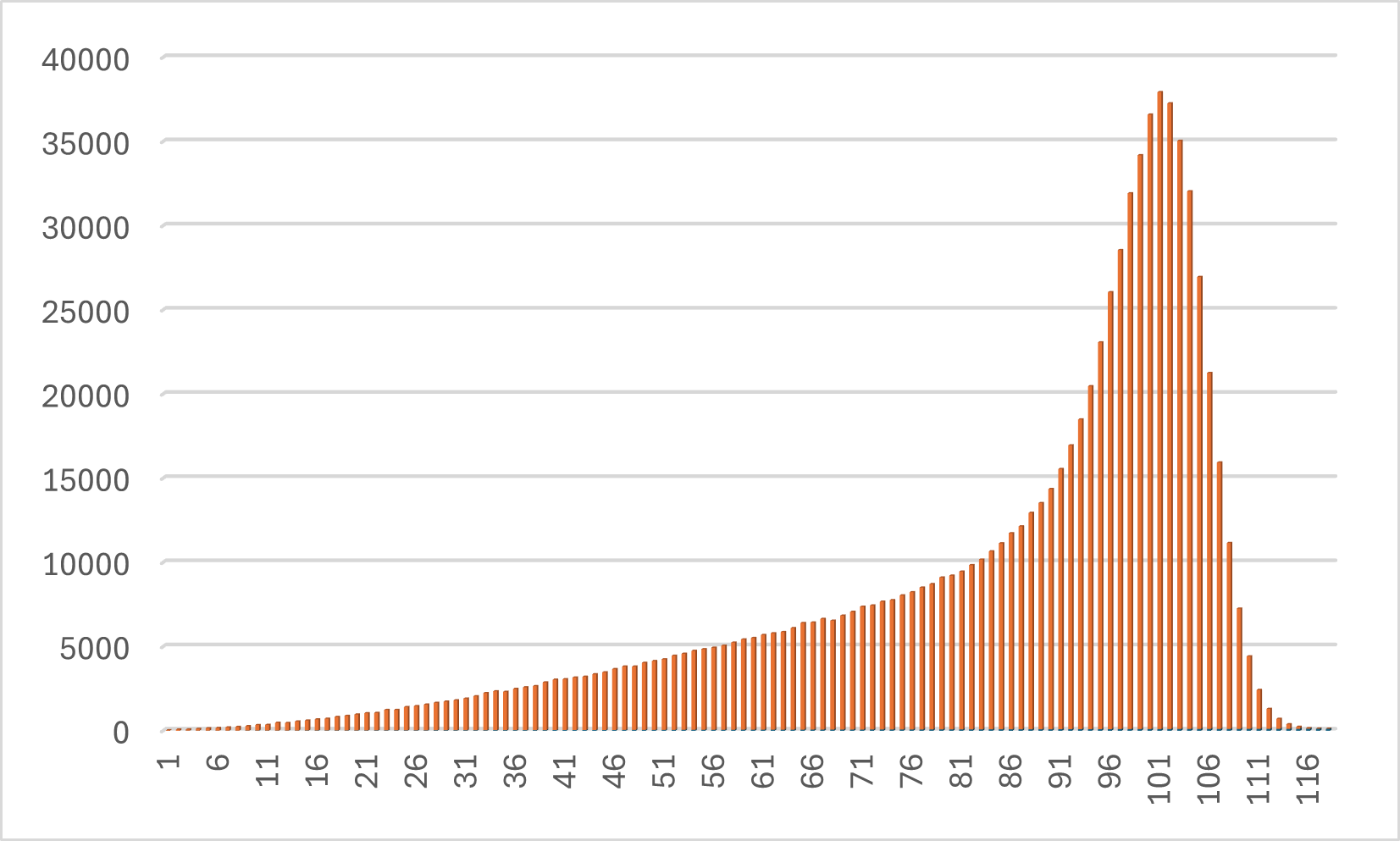}
    \caption{Amount of neighbors at a distance of 5 grid ``steps'' (counted only in 2D plane of individual layers). Only \revnr{3.2~\%} of all voxels have less than \revnr{33} neighbors. }
    \label{fig:ASTM_Stat_NeighboursNrHist_Dist5}
\end{figure}

The ``false positive'' recognition of the voxels appears as a ``halo'' around the reconstructed object, as illustrated in Figure~\ref{fig:ASTM2_Rasterized_zoom-in_halo}.
The Voxel Pruning (described in Section~\ref{sec:reconstruction:voxel_pruning}) is the approach we chose to reduce amount of such false positives.
To determine the parameters for pruning based on neighborhood relationships, we collected several statistics.
The length of the ``gap'' stretches (the histogram is shown in Figure~\ref{fig:ASTM_Stat_GapStretchHist}) informed our decision on the neighborhood range. 
The distribution of the number of neighbors across all voxels (the histogram is shown in Figure~\ref{fig:ASTM_Stat_NeighboursNrHist_Dist5}) informed our decision on the minimum number of neighbors required for a voxel not to be removed.


\begin{figure}
  \centering
  \includegraphics[width=0.7\linewidth]{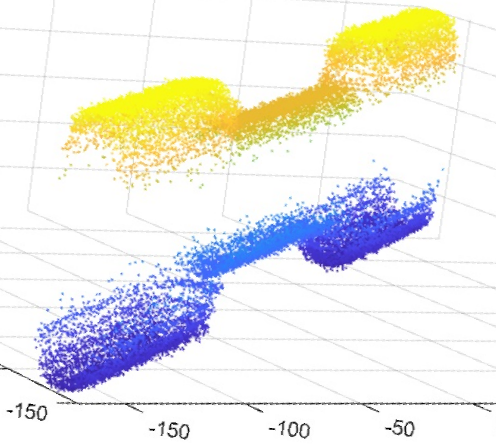}
  \caption{Reconstructed ASTM E2 model where each voxel below middle layer (51) is projected ``down'' and all voxels in layers above are projected ``up'' to the single voxel with the same ${(x, y)}$ and either smallest or largest $z$ coordinates. 
  The ${HitCtr}$ is aggregated across all the projected voxels with the same ${(x, y)}$ and then pruned with the minimum ${HitCtr}$ threshold at 20. 
  \todo{replace with image of pruning by min. 55, as used in code!} 
  }
  
  \label{fig:ASTM2_Rasterized_Projected}
\end{figure}


\section{Matters of Gap Filling}
\label{app:gap_filling}

For Gap Filling (described in Section~\ref{sec:reconstruction:gap_filling}), we projected all voxels on the terminal layer(s).
For the Gear model, we displayed the projection in the section.
The projection for ASTM E8 model is more complex, as it is projected both up and down (see Figure~\ref{fig:ASTM2_Rasterized_Projected}).


\section{Determining Distortion}
\label{app:determining_distortion}

\begin{figure}
  \centering
  \includegraphics[width=0.65\linewidth]{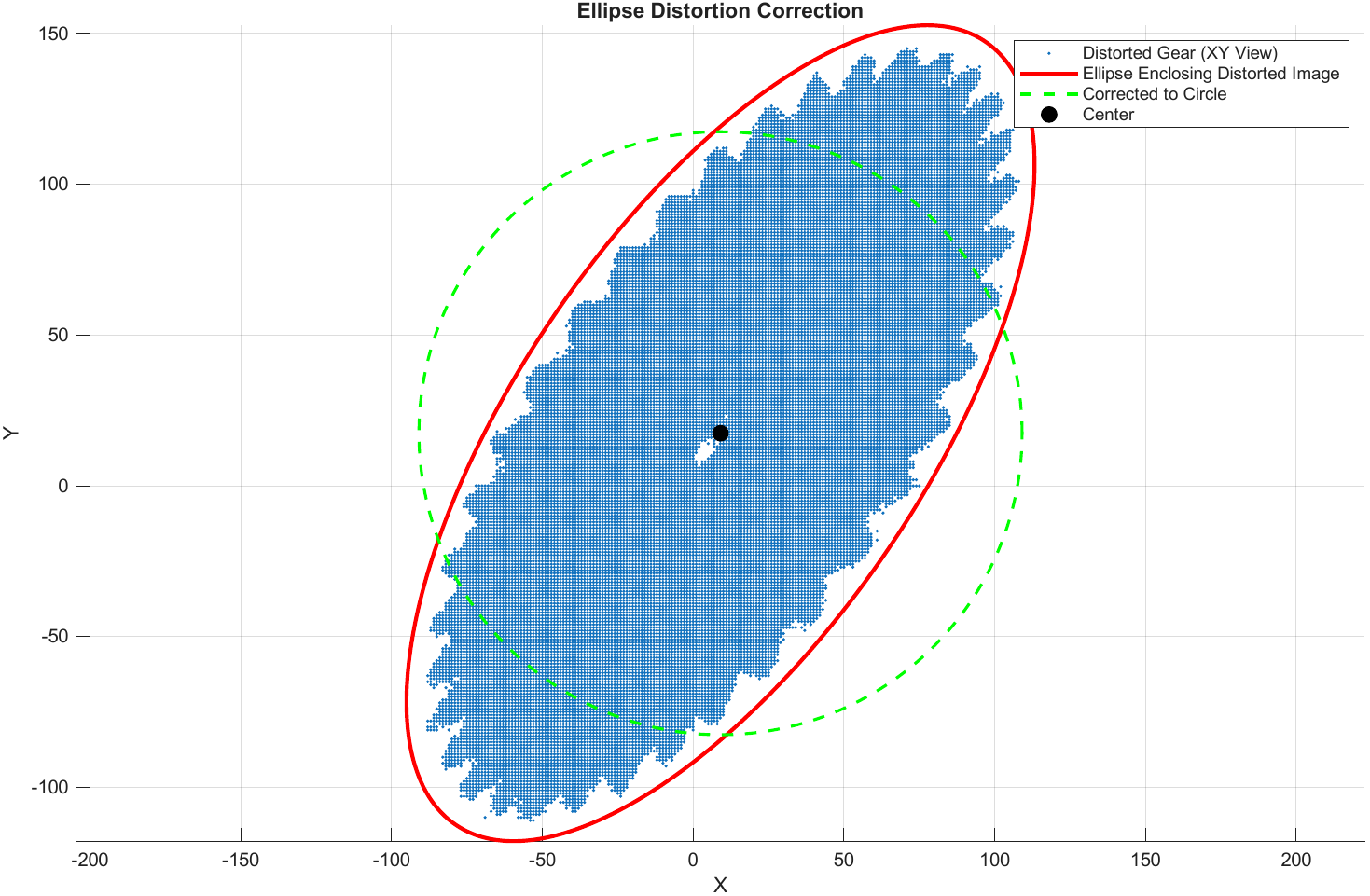}
  \caption{Fitting ellipse around ``collapsed'' Gear reconstruction and the reference circle for XY distortion correction.}
  
  \label{fig:object_reconstruction_NO_pruning}
\end{figure}

To determine distortion in the XY plane (discussed in Section~\ref{sec:reconstruction:distortion_correction}), we used the fitting ellipse around the projected Gear reconstructed model and the reference circle, as shown in Figure~\ref{fig:object_reconstruction_NO_pruning}.

\section{Major Stages of Reconstruction -- Illustration}
\label{app:reconstruction_stages}

\begin{figure}[htbp]
    \centering
    
    \begin{subfigure}[b]{0.95\linewidth}
        \centering
		\includegraphics[width=\linewidth]{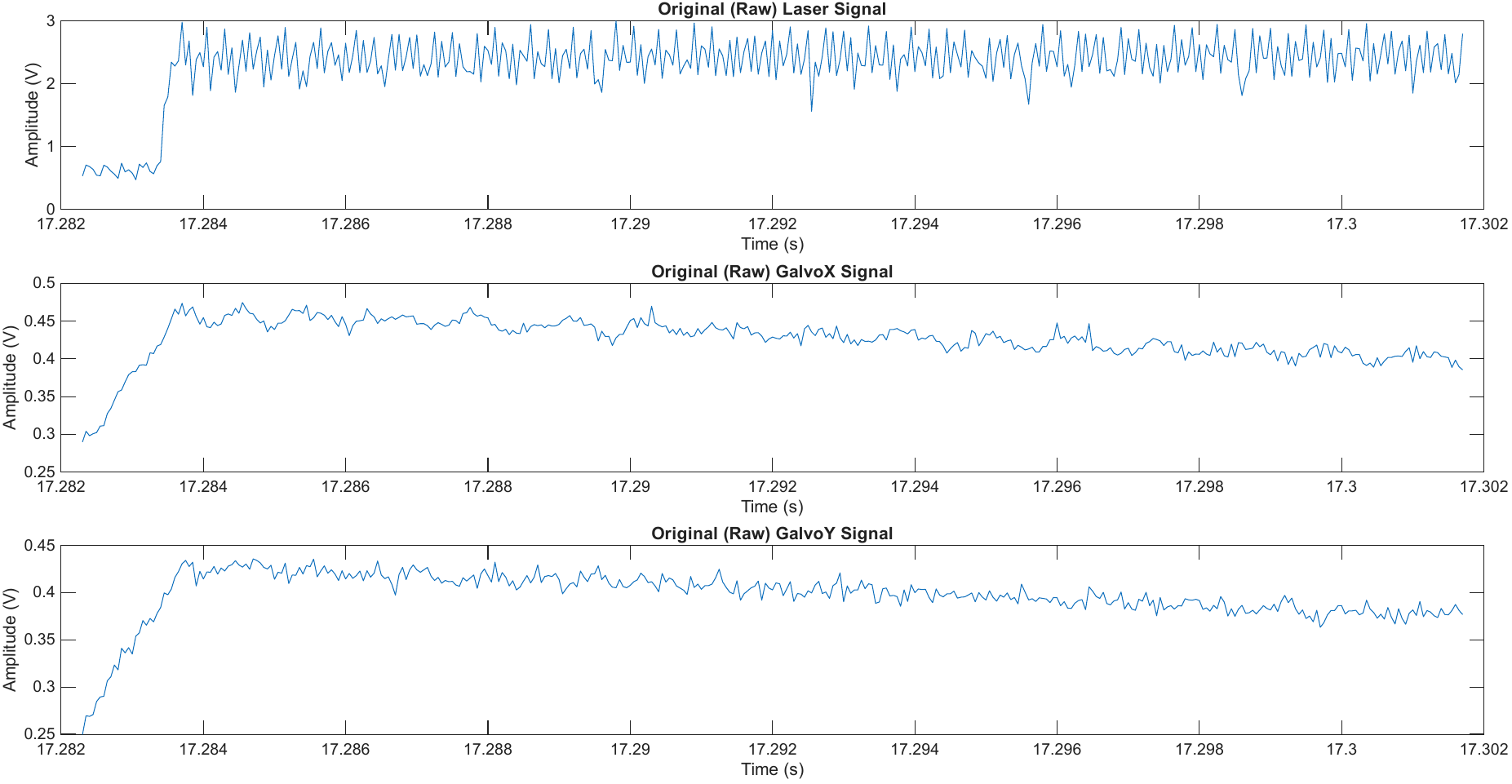}
        \caption{Raw signals}
		\label{fig:LaserGalvoXY_raw}
    \end{subfigure}%
    
    \begin{subfigure}[b]{0.95\linewidth}
        \centering
		\includegraphics[width=\linewidth]{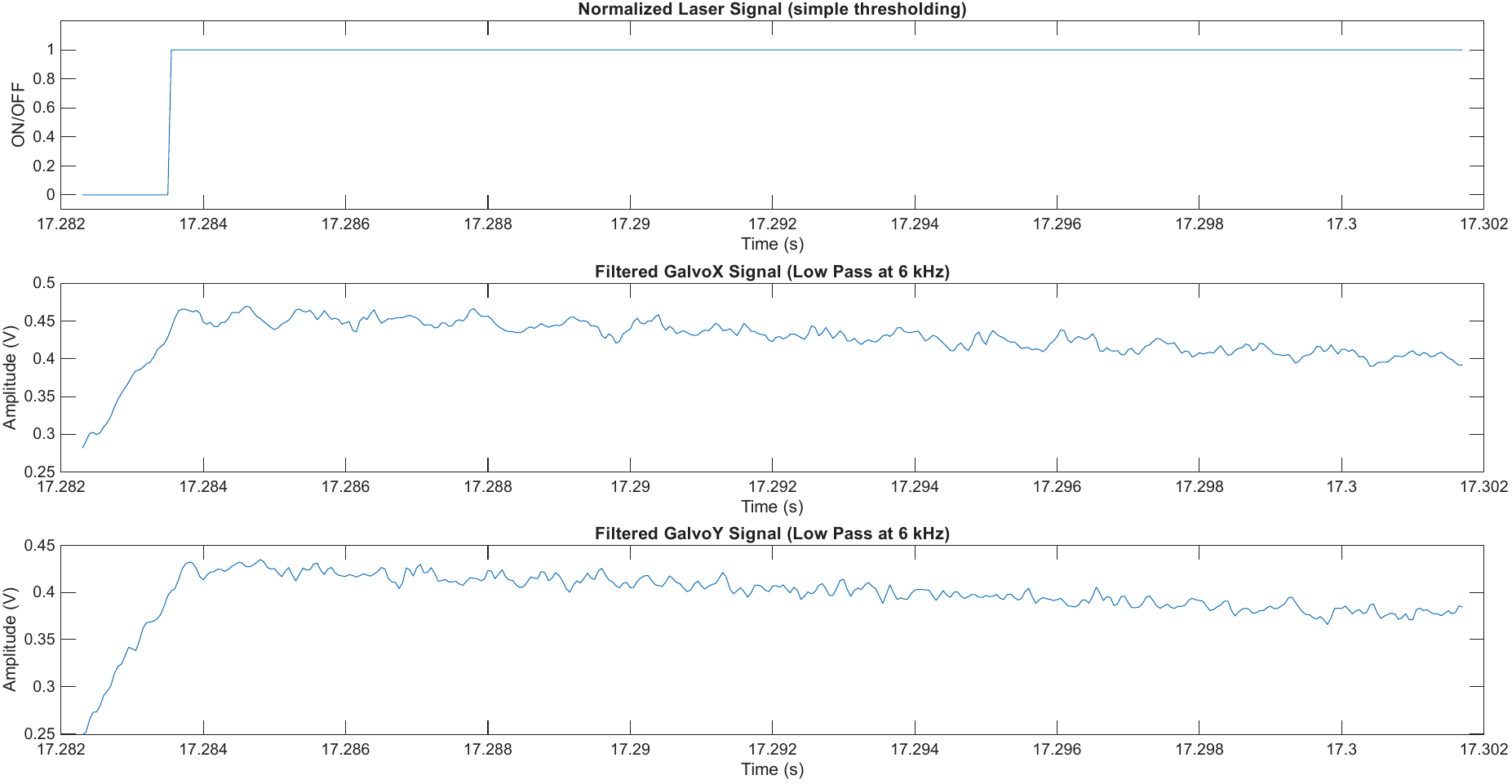}
        \caption{Preprocessed signals}
		\label{fig:LaserGalvoXY_preprocessed}
    \end{subfigure}%

	\caption{A section of raw traces and the corresponding signals after post-processing. Laser was normalized using algorithm shown in Figure~\ref{fig:code:laser_normalize}; the high frequency noise has been removed from galvanometer signals using 4th order Butterworth low pass filter with the \revnr{6 kHz} cutoff frequencies. The signals are for a capture of ASTM E8 print, selection between polling indices \revnr{345646 to 345934} (diagram shows conversion to seconds under 20 kHz poling frequency). }
	\label{fig:app:signal_preprocessing}
\end{figure}

The very first step is preprocessing of the raw power side-channel signals (described in Section~\ref{sec:reconstruction:filtering}).
Figure~\ref{fig:app:signal_preprocessing} shows a short section of the signal before and after this step.

\begin{figure}[htbp!]
  \centering
  \includegraphics[width=0.9\linewidth]{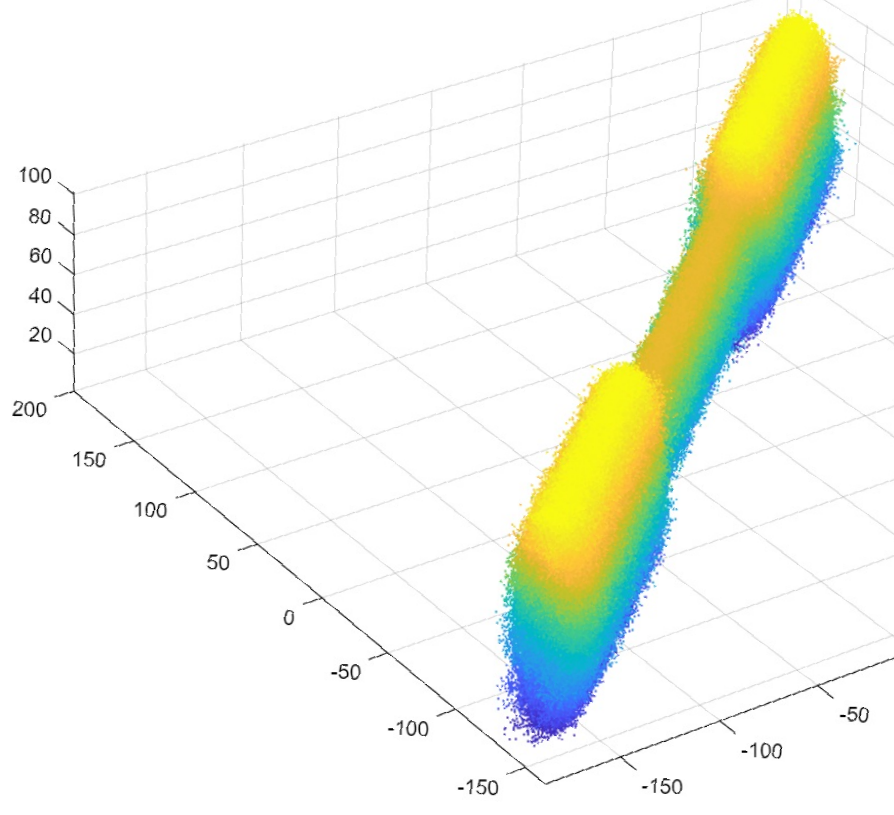}
  \caption{Rasterizationn of specimens at \revnr{0.0025~V} for GalvoX and GalvoY. Voxelization with \revnr{+1} of height for every Layer.}
  \label{fig:ASTM_rasterized}
\end{figure}

The next notable stage of reconstruction is the rasterization (described in Section~\ref{sec:reconstruction:rasterization}).
An example of rasterized ASTM E8 samples is shown in Figure~\ref{fig:ASTM_rasterized}. The choice of raster size caused distortion along Z-axis when point cloud is displyed in (RasterX, RasterY, LayerNr) coordinates.

\begin{figure}[htbp!]
  \centering
  \includegraphics[width=0.45\linewidth]{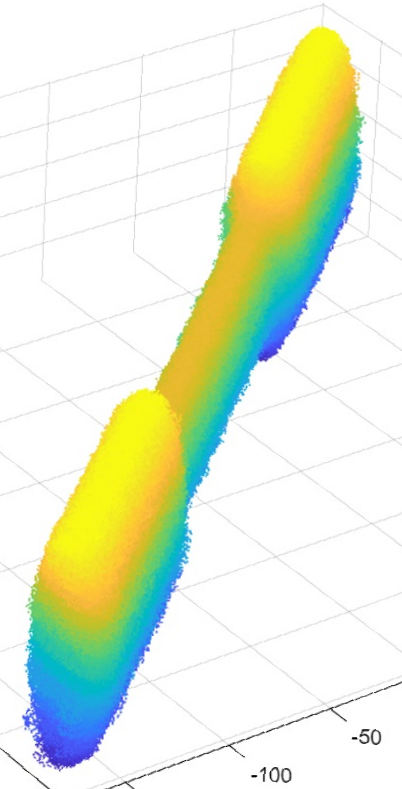}
  \caption{Pruned reconstructed model (from a single trace). The pruning conditions were that minimum $HitCtr$ is 1, and the amount of neighbors in the 2D range of 5 should be at least 33. Even though some false positives remain, the amount of spurious points in ``halo'' is significantly reduced. }
  \label{fig:app:voxel_pruning}
\end{figure}

The next stage is the Voxel Pruning, as described in Section~\ref{sec:reconstruction:voxel_pruning}.
After the pruning stage. a significant reconstruction of the False Positive voxels ``halo'' around the recovered model is achieved (see Figure~\ref{fig:app:voxel_pruning}).

\begin{figure}[htbp!]
  \centering
  \includegraphics[width=0.95\linewidth]{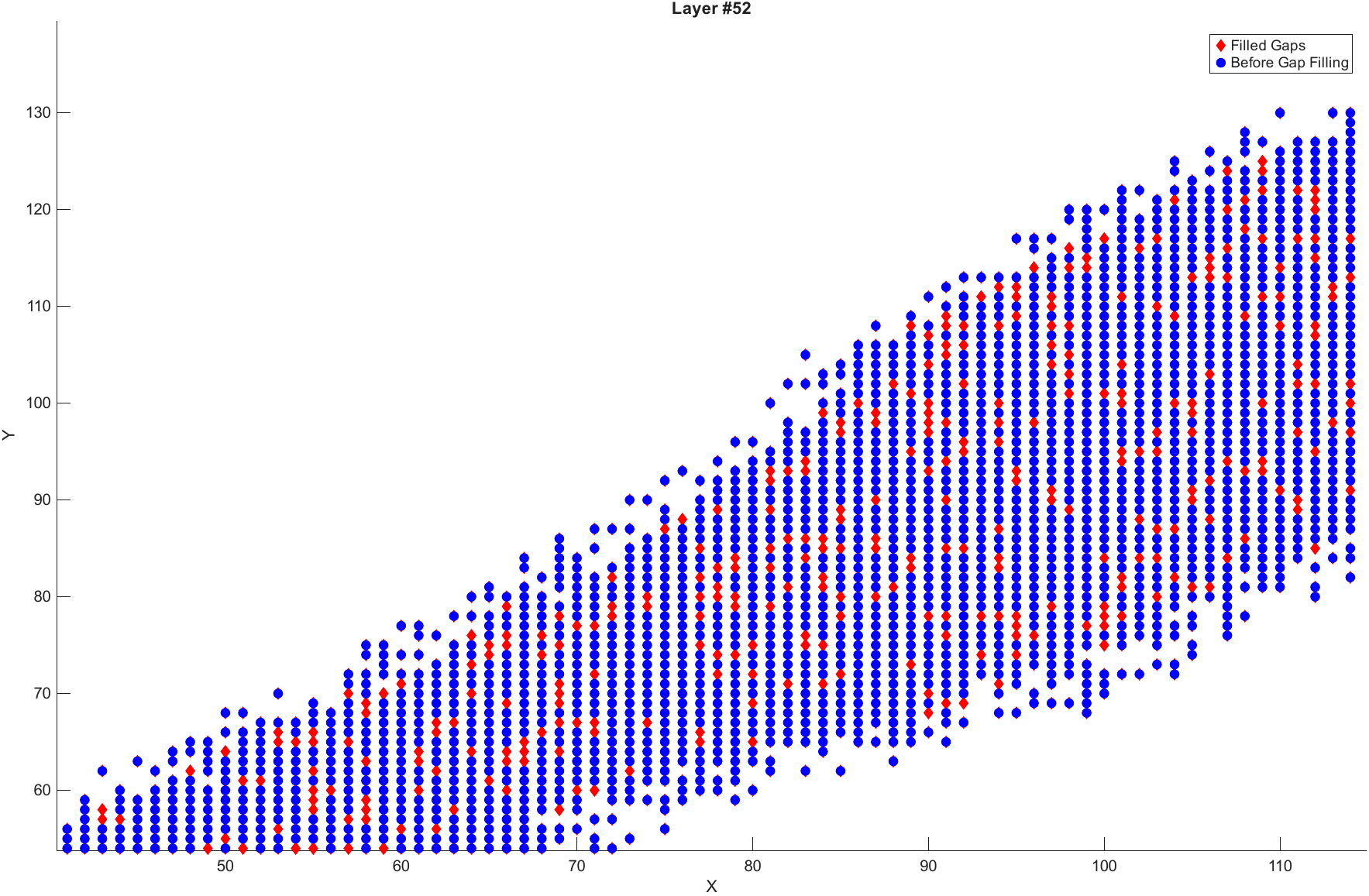}
  \caption{Example of Gap Filling. Blue dots indicate original voxels, red dots indicate voxels added to fill identified gaps. Note that not all identified gaps also qualify for being filled. }
  \label{fig:app:gap_filling}
\end{figure}

\newpage 
The gaps are filled according to the approach described in Section~\ref{sec:reconstruction:gap_filling}.
Figure~\ref{fig:app:gap_filling} illustrates a ``zoom-in'' at a single layer of reconstructed Gear model before and after the procedure.

\begin{figure}[htbp]
    \centering
    
    \begin{subfigure}[b]{0.45\linewidth}
        \centering
		\includegraphics[width=\linewidth]{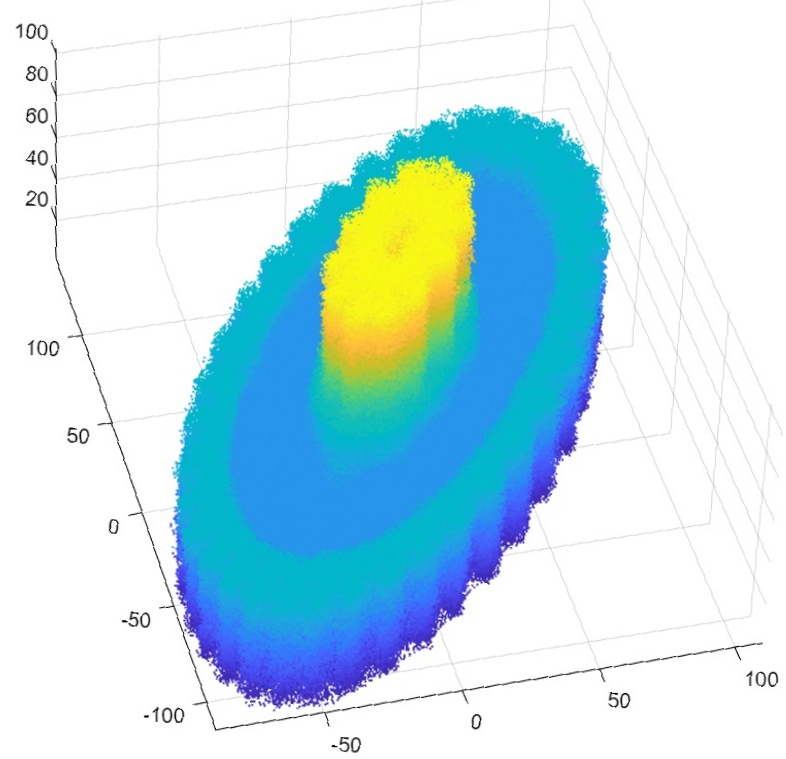}
        \caption{Before XY Correction}
		\label{fig:DistortionCorrectionXY_Gear1_3D_Before}
    \end{subfigure}%
    ~
    \begin{subfigure}[b]{0.45\linewidth}
        \centering
		\includegraphics[width=\linewidth]{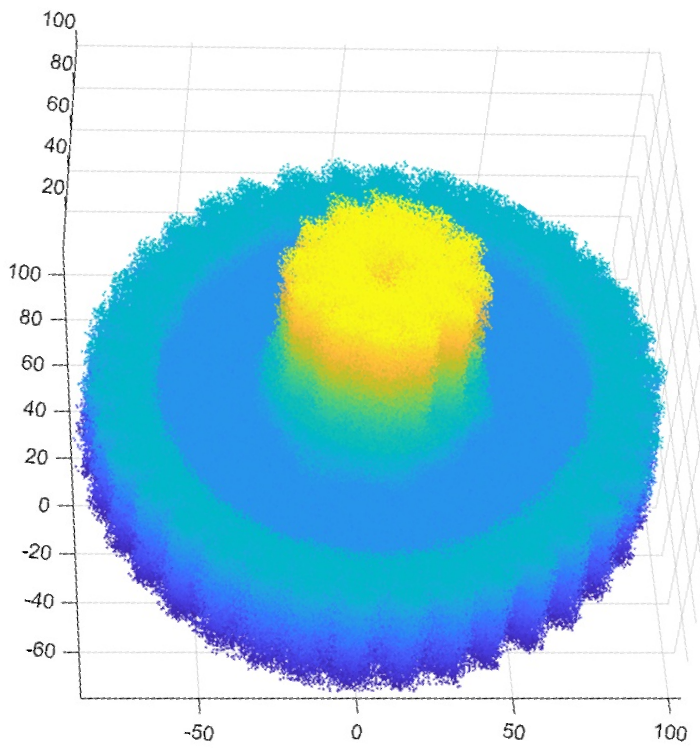}
        \caption{After XY Correction}
		\label{fig:DistortionCorrectionXY_Gear1_3D_After}
    \end{subfigure}%

	\caption{Illustration of Distortion Correction in XY Plane. }
	\label{fig:app:distortion_correction_xy}
\end{figure}

\begin{figure}[htbp]
    \centering
    
    \begin{subfigure}[b]{0.45\linewidth}
        \centering
		\includegraphics[width=\linewidth]{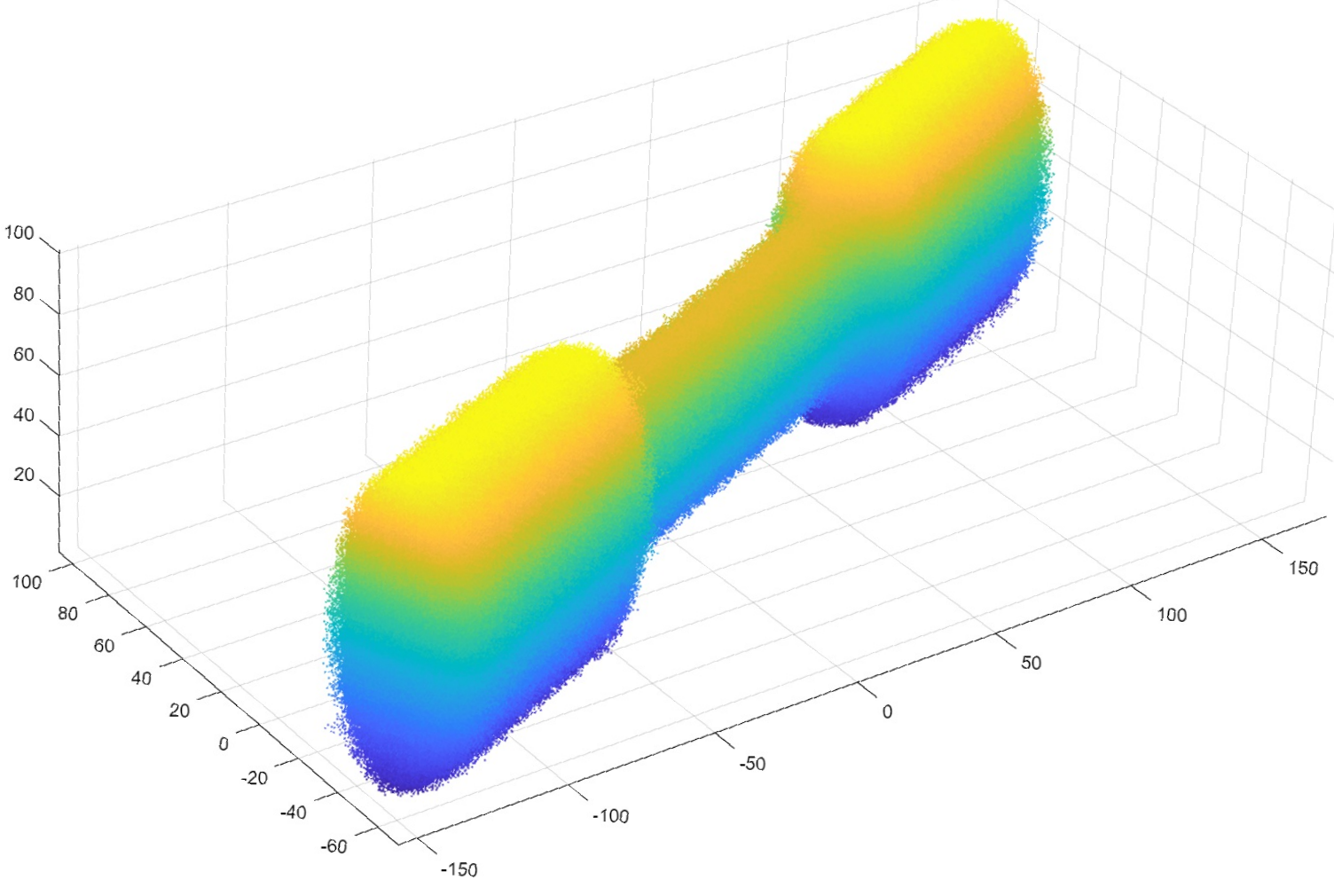}
        \caption{Before Z Correction}
		\label{fig:DistortionCorrectionZ_ASTM2_3D_Before}
    \end{subfigure}%
    ~
    \begin{subfigure}[b]{0.45\linewidth}
        \centering
		\includegraphics[width=\linewidth]{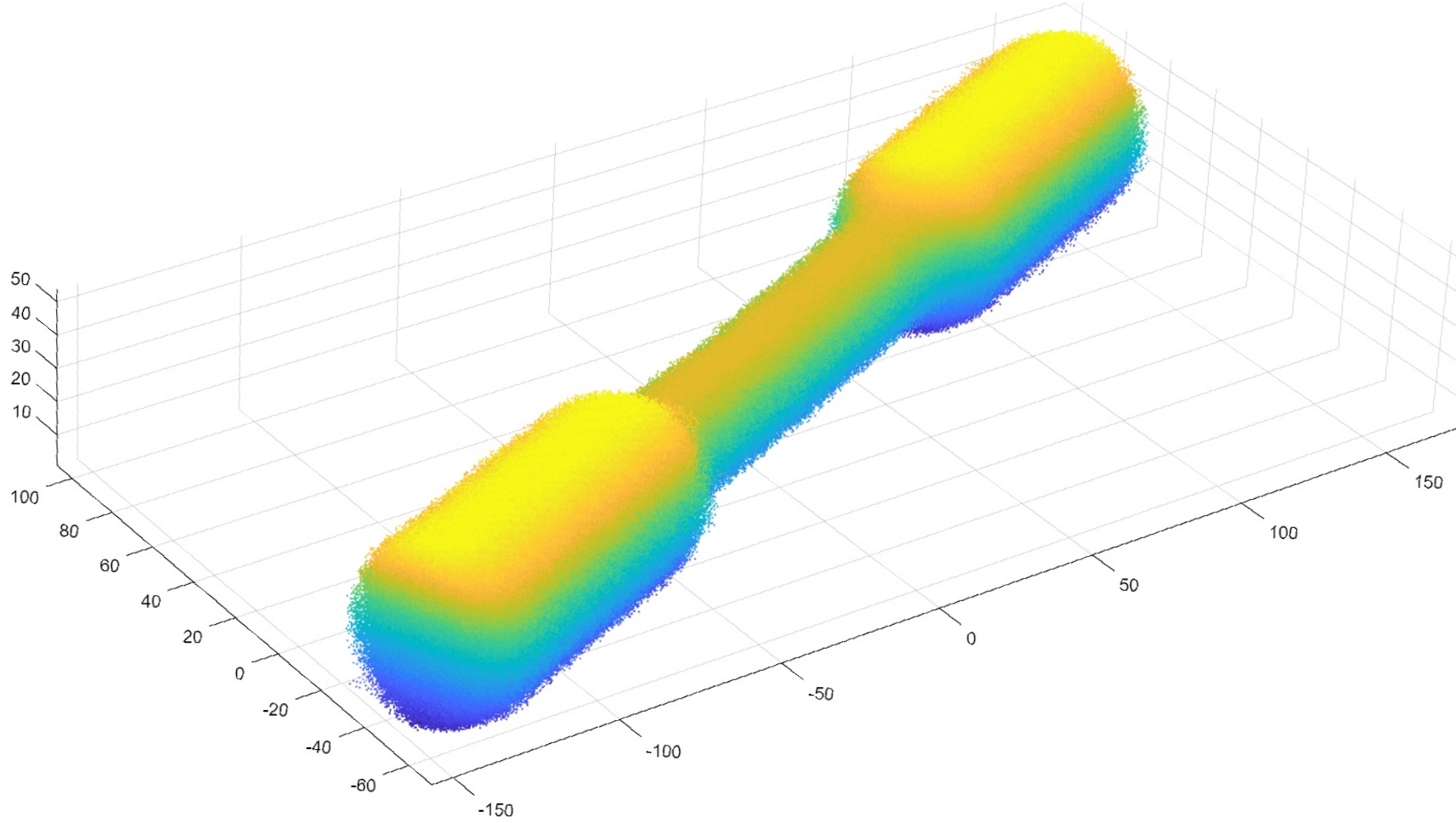}
        \caption{After Z Correction}
		\label{fig:DistortionCorrectionZ_ASTM2_3D_After}
    \end{subfigure}%

	\caption{Illustration of Distortion Correction along Z-axis. }
	\label{fig:app:distortion_correction_z}
\end{figure}

The next step is to correct for distortions, as described in Section~\ref{sec:reconstruction:distortion_correction}.
The correction in the XY plane is better illustrated in the Gear model and along the Z-axis in the ASTM E8 specimen (see Figures~\ref{fig:app:distortion_correction_xy} and ~\ref{fig:app:distortion_correction_z}, respectively).

\section{Comparison - Reconstructed to STL Model}
\label{app:comparison}

\begin{figure}[htbp]
    \centering
    
    \begin{subfigure}[b]{0.95\linewidth}
        \centering
		\includegraphics[width=\linewidth]{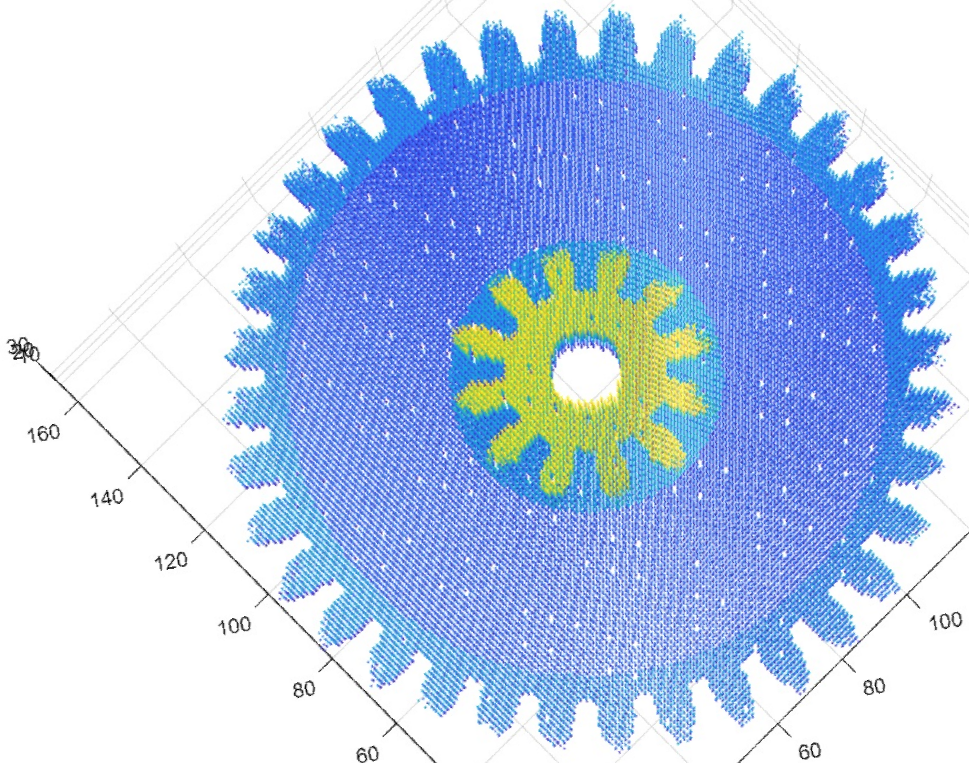}
        \caption{True Positives}
		\label{fig:Comparison_REC2STL_3D_Gear_TP}
    \end{subfigure}%
    
    \begin{subfigure}[b]{0.95\linewidth}
        \centering
		\includegraphics[width=\linewidth]{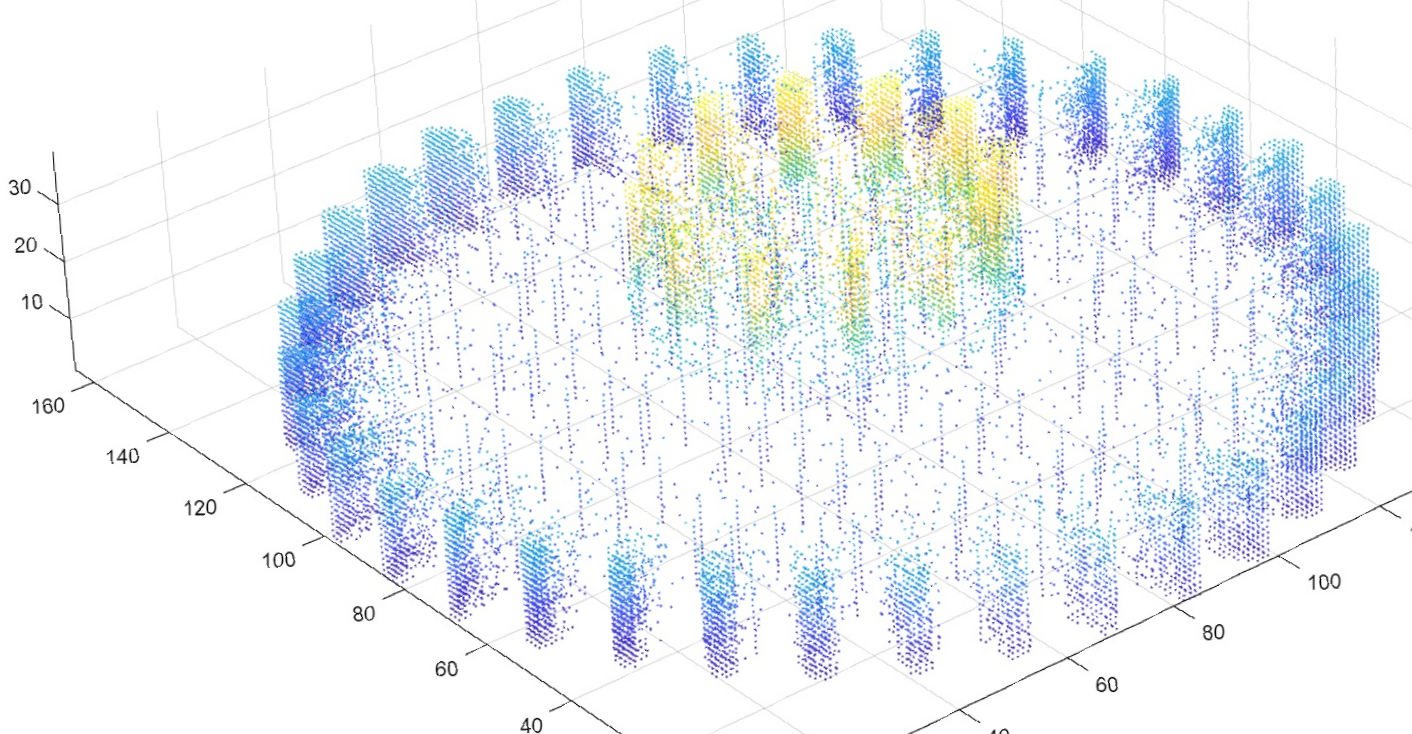}
        \caption{False Negatives}
		\label{fig:Comparison_REC2STL_3D_Gear_FN}
    \end{subfigure}%

    \begin{subfigure}[b]{0.95\linewidth}
        \centering
		\includegraphics[width=\linewidth]{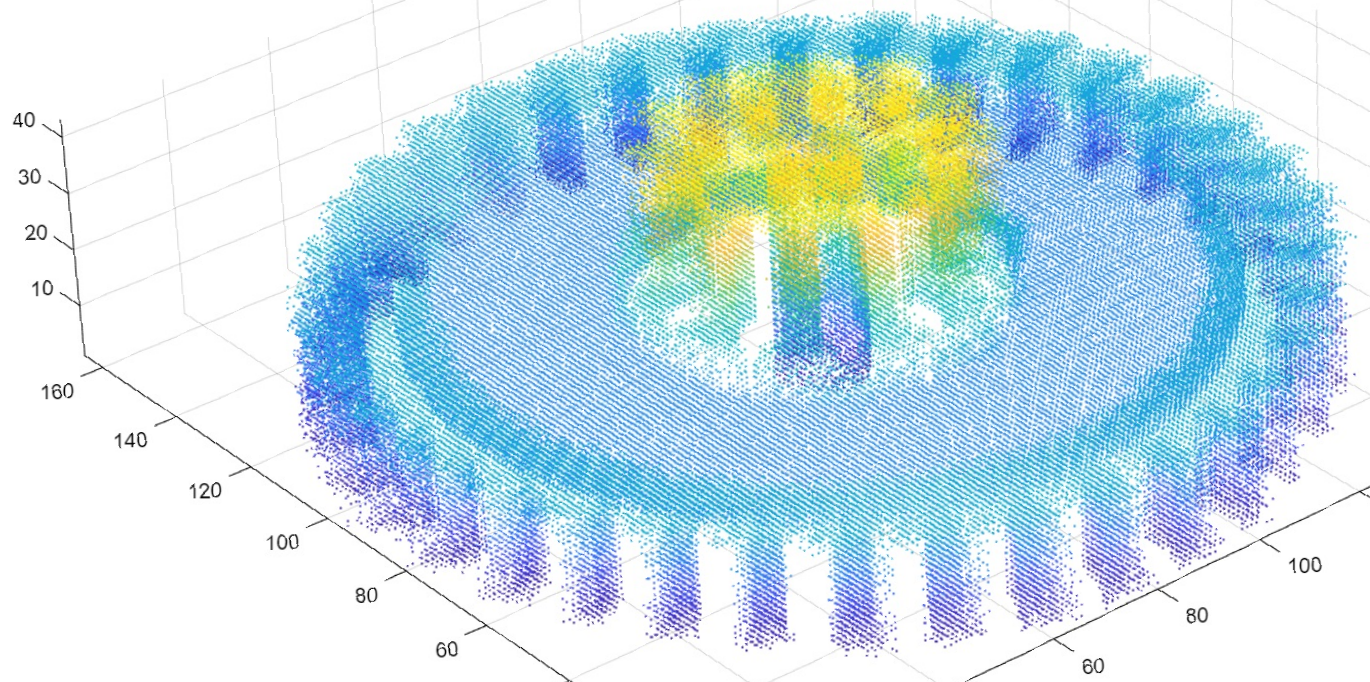}
        \caption{False Positives}
		\label{fig:Comparison_REC2STL_3D_Gear_FP}
    \end{subfigure}%

	\caption{Point Cloud visualizing results of comparison between Reconstructed and the original STL model.}
	\label{fig:app:comparison_REC2STL}
\end{figure}

During the comparison between the original STL and the reconstructed model, we saved point clouds of True Positives, True Negative, and False Positive voxels.
There results are demonstrated in Figure~\ref{fig:app:comparison_REC2STL}.

%% file: TDTonPBF_0_main.bbl
\begin{thebibliography}{10}
\providecommand{\url}[1]{#1}
\csname url@samestyle\endcsname
\providecommand{\newblock}{\relax}
\providecommand{\bibinfo}[2]{#2}
\providecommand{\BIBentrySTDinterwordspacing}{\spaceskip=0pt\relax}
\providecommand{\BIBentryALTinterwordstretchfactor}{4}
\providecommand{\BIBentryALTinterwordspacing}{\spaceskip=\fontdimen2\font plus
\BIBentryALTinterwordstretchfactor\fontdimen3\font minus \fontdimen4\font\relax}
\providecommand{\BIBforeignlanguage}[2]{{%
\expandafter\ifx\csname l@#1\endcsname\relax
\typeout{** WARNING: IEEEtran.bst: No hyphenation pattern has been}%
\typeout{** loaded for the language `#1'. Using the pattern for}%
\typeout{** the default language instead.}%
\else
\language=\csname l@#1\endcsname
\fi
#2}}
\providecommand{\BIBdecl}{\relax}
\BIBdecl

\bibitem{astmF2792}
ASTM, \emph{ASTM F2792-12a, Standard Terminology for Additive Manufacturing Technologies (Withdrawn)}.\hskip 1em plus 0.5em minus 0.4em\relax ASTM, 2015.

\bibitem{wohlers2025report}
\BIBentryALTinterwordspacing
W.~Associates and A.~International, \emph{Wohlers Report 2025: 3D Printing and Additive Manufacturing State of the Industry}.\hskip 1em plus 0.5em minus 0.4em\relax Fort Collins, CO, USA: Wohlers Associates, 2025, 30th edition; over 230 contributors across six continents. [Online]. Available: \url{https://wohlersassociates.com/product/wr2025/}
\BIBentrySTDinterwordspacing

\bibitem{yampolskiy2022state}
M.~Yampolskiy, P.~Bates, M.~Seifi, and N.~Shamsaei, ``State of security awareness in the am industry: 2020 survey,'' \emph{arXiv preprint arXiv:2209.03073}, 2022.

\bibitem{url20243dprintingbusiness}
``{3D Printing Business Directory},'' https://www.voxelmatters.directory/, 2024, [Online; accessed 7 May 2024].

\bibitem{wade2016digital}
T.~J. Wade and S.~Subramanya, ``Digital rights management in 3d printing: A proposed reference architecture for design-to-fabrication security and licensing,'' in \emph{International Conference on Breakthrough in Engineering, Science \& Technology—2016 (INC-BEST'16)}, 2016.

\bibitem{alkaabi2020blockchain}
N.~Alkaabi, K.~Salah, R.~Jayaraman, J.~Arshad, M.~Omar \emph{et~al.}, ``Blockchain-based traceability and management for additive manufacturing,'' \emph{IEEE access}, vol.~8, pp. 188\,363--188\,377, 2020.

\bibitem{baumann2017model}
F.~Baumann, T.~Ludwig, N.~Darwin~Abele, S.~Hoffmann, and D.~Roller, ``Model-data streaming for additive manufacturing securing intellectual property,'' \emph{Smart and Sustainable Manufacturing Systems}, vol.~1, no.~1, pp. 142--152, 2017.

\bibitem{kolter2025streaming}
M.~Kolter, S.~Dirks, S.~Scheres, and J.~H. Schleifenbaum, ``Streaming in metal additive manufacturing: a catalyst for secure distributed manufacturing,'' \emph{International Journal of Production Research}, pp. 1--15, 2025.

\bibitem{kok2022design}
X.~W. Kok, A.~Singh, and B.~T. Raimi-Abraham, ``A design approach to optimise secure remote three-dimensional (3d) printing: A proof-of-concept study towards advancement in telemedicine,'' in \emph{Healthcare}, vol.~10, no.~6.\hskip 1em plus 0.5em minus 0.4em\relax MDPI, 2022, p. 1114.

\bibitem{safford2019hardware}
D.~R. Safford and M.~Wiseman, ``Hardware rooted trust for additive manufacturing,'' \emph{IEEE Access}, vol.~7, pp. 79\,211--79\,215, 2019.

\bibitem{cultice2023novel}
T.~Cultice, J.~Clark, W.~Yang, and H.~Thapliyal, ``A novel hierarchical security solution for controller-area-network-based 3d printing in a post-quantum world,'' \emph{Sensors}, vol.~23, no.~24, p. 9886, 2023.

\bibitem{identify2020info}
Identify3D, ``Identify3d info sheet,'' \url{https://identify3d.com/wp-content/uploads/2019/06/Identify3DInfosheet-1.pdf}, 2022.

\bibitem{materialise2022pressrelease}
Materialise, ``Materialise makes co-am the most secure platform for distributed manufacturing with acquisition of identify3d,'' \url{https://www.materialise.com/en/news/press-releases/acquisition-identify3d}, 2022.

\bibitem{assembrix2023web}
Assembrix, ``Assembrix website,'' \url{https://assembrix.com/}, 2023.

\bibitem{collberg2009surreptitious}
C.~Collberg and J.~Nagra, \emph{Surreptitious Software: Obfuscation, Watermarking, and Tamperproofing for Software Protection}.\hskip 1em plus 0.5em minus 0.4em\relax Addison-Wesley Professional, 2009.

\bibitem{akhunzada2015man}
A.~Akhunzada, M.~Sookhak, N.~B. Anuar, A.~Gani, E.~Ahmed, M.~Shiraz, S.~Furnell, A.~Hayat, and M.~K. Khan, ``Man-at-the-end attacks: Analysis, taxonomy, human aspects, motivation and future directions,'' \emph{Journal of Network and Computer Applications}, vol.~48, pp. 44--57, 2015.

\bibitem{yampolskiy2014intellectual}
M.~Yampolskiy, T.~R. Andel, J.~T. McDonald, W.~B. Glisson, and A.~Yasinsac, ``Intellectual property protection in additive layer manufacturing: Requirements for secure outsourcing,'' in \emph{Proceedings of the 4th Program Protection and Reverse Engineering Workshop}, 2014, pp. 1--9.

\bibitem{faruque2016acoustic}
M.~A. Al~Faruque, S.~R. Chhetri, A.~Canedo, and J.~Wan, ``Acoustic side-channel attacks on additive manufacturing systems,'' in \emph{2016 ACM/IEEE 7th international conference on Cyber-Physical Systems (ICCPS)}.\hskip 1em plus 0.5em minus 0.4em\relax IEEE, 2016, pp. 1--10.

\bibitem{hojjati2016leave}
A.~Hojjati, A.~Adhikari, K.~Struckmann, E.~Chou, T.~N. Tho~Nguyen, K.~Madan, M.~S. Winslett, C.~A. Gunter, and W.~P. King, ``{Leave your Phone at the Door: Side Channels that Reveal Factory Floor Secrets},'' in \emph{Proceedings of the 2016 ACM SIGSAC Conference on Computer and Communications Security}, 2016, pp. 883--894.

\bibitem{song2016my}
C.~Song, F.~Lin, Z.~Ba, K.~Ren, C.~Zhou, and W.~Xu, ``{My Smartphone Knows What You Print: Exploring Smartphone-Based Side-Channel Attacks against 3D Printers},'' in \emph{Proceedings of the 2016 ACM SIGSAC Conference on Computer and Communications Security}, 2016, pp. 895--907.

\bibitem{gatlin2021encryption}
J.~Gatlin, S.~Belikovetsky, Y.~Elovici, A.~Skjellum, J.~Lubell, P.~Witherell, and M.~Yampolskiy, ``{Encryption is Futile: Reconstructing 3D-Printed Models using the Power Side-Channel},'' in \emph{Proceedings of the 24th International Symposium on Research in Attacks, Intrusions and Defenses}, 2021, pp. 135--147.

\bibitem{pearce2022flaw3d}
H.~Pearce, K.~Yanamandra, N.~Gupta, and R.~Karri, ``Flaw3d: A trojan-based cyber attack on the physical outcomes of additive manufacturing,'' \emph{IEEE/ASME Transactions on Mechatronics}, vol.~27, no.~6, pp. 5361--5370, 2022.

\bibitem{randolph2020power}
M.~Randolph and W.~Diehl, ``Power side-channel attack analysis: A review of 20 years of study for the layman,'' \emph{Cryptography}, vol.~4, no.~2, p.~15, 2020.

\bibitem{sayakkara2019survey}
A.~Sayakkara, N.-A. Le-Khac, and M.~Scanlon, ``A survey of electromagnetic side-channel attacks and discussion on their case-progressing potential for digital forensics,'' \emph{Digital Investigation}, vol.~29, pp. 43--54, 2019.

\bibitem{ge2018survey}
Q.~Ge, Y.~Yarom, D.~Cock, and G.~Heiser, ``A survey of microarchitectural timing attacks and countermeasures on contemporary hardware,'' \emph{Journal of Cryptographic Engineering}, vol.~8, no.~1, pp. 1--27, 2018.

\bibitem{kocher1999differential}
P.~Kocher, J.~Jaffe, and B.~Jun, ``{Differential Power Analysis},'' in \emph{Annual International Cryptology Conference}.\hskip 1em plus 0.5em minus 0.4em\relax Springer, 1999, pp. 388--397.

\bibitem{ravi2004security}
S.~Ravi, A.~Raghunathan, P.~Kocher, and S.~Hattangady, ``Security in embedded systems: Design challenges,'' \emph{ACM Transactions on Embedded Computing Systems (TECS)}, vol.~3, no.~3, pp. 461--491, 2004.

\bibitem{mangard2007power}
S.~Mangard, E.~Oswald, and T.~Popp, \emph{Power analysis attacks: Revealing the secrets of smart cards}.\hskip 1em plus 0.5em minus 0.4em\relax Springer, 2007.

\bibitem{rankl2003overview}
W.~Rankl, ``Overview about attacks on smart cards,'' \emph{Information Security Technical Report}, vol.~8, no.~1, pp. 67--84, 2003.

\bibitem{standaert2009unified}
F.-X. Standaert, T.~G. Malkin, and M.~Yung, ``A unified framework for the analysis of side-channel key recovery attacks,'' in \emph{Annual international conference on the theory and applications of cryptographic techniques}.\hskip 1em plus 0.5em minus 0.4em\relax Springer, 2009, pp. 443--461.

\bibitem{devi2020side}
M.~Devi and A.~Majumder, ``Side-channel attack in internet of things: A survey,'' in \emph{Applications of Internet of Things: Proceedings of ICCCIOT 2020}.\hskip 1em plus 0.5em minus 0.4em\relax Springer, 2020, pp. 213--222.

\bibitem{zankl2018side}
A.~Zankl, H.~Seuschek, G.~Irazoqui, and B.~Gulmezoglu, ``Side-channel attacks in the internet of things: threats and challenges,'' in \emph{Solutions for Cyber-Physical Systems Ubiquity}.\hskip 1em plus 0.5em minus 0.4em\relax IGI Global, 2018, pp. 325--357.

\bibitem{quinn2009privacy}
E.~L. Quinn, ``Privacy and the new energy infrastructure,'' \emph{Available at SSRN 1370731}, 2009.

\bibitem{ristenpart2009hey}
T.~Ristenpart, E.~Tromer, H.~Shacham, and S.~Savage, ``Hey, you, get off of my cloud: exploring information leakage in third-party compute clouds,'' in \emph{Proceedings of the 16th ACM conference on Computer and communications security}, 2009, pp. 199--212.

\bibitem{zhang2012cross}
Y.~Zhang, A.~Juels, M.~K. Reiter, and T.~Ristenpart, ``Cross-vm side channels and their use to extract private keys,'' in \emph{Proceedings of the 2012 ACM conference on Computer and communications security}, 2012, pp. 305--316.

\bibitem{lipp2018meltdown}
M.~Lipp, M.~Schwarz, D.~Gruss, T.~Prescher, W.~Haas, S.~Mangard, P.~Kocher, D.~Genkin, Y.~Yarom, and M.~Hamburg, ``Meltdown,'' \emph{arXiv preprint arXiv:1801.01207}, 2018.

\bibitem{kocher2020spectre}
P.~Kocher, J.~Horn, A.~Fogh, D.~Genkin, D.~Gruss, W.~Haas, M.~Hamburg, M.~Lipp, S.~Mangard, T.~Prescher \emph{et~al.}, ``Spectre attacks: Exploiting speculative execution,'' \emph{Communications of the ACM}, vol.~63, no.~7, pp. 93--101, 2020.

\bibitem{koruyeh2018spectre}
E.~M. Koruyeh, K.~N. Khasawneh, C.~Song, and N.~Abu-Ghazaleh, ``Spectre returns! speculation attacks using the return stack buffer,'' in \emph{12th USENIX Workshop on Offensive Technologies (WOOT 18)}, 2018.

\bibitem{agrawal2007trojan}
D.~Agrawal, S.~Baktir, D.~Karakoyunlu, P.~Rohatgi, and B.~Sunar, ``Trojan detection using ic fingerprinting,'' in \emph{Security and Privacy, 2007. SP'07. IEEE Symposium on}.\hskip 1em plus 0.5em minus 0.4em\relax IEEE, 2007, pp. 296--310.

\bibitem{narasimhan2012hardware}
S.~Narasimhan, D.~Du, R.~S. Chakraborty, S.~Paul, F.~G. Wolff, C.~A. Papachristou, K.~Roy, and S.~Bhunia, ``Hardware trojan detection by multiple-parameter side-channel analysis,'' \emph{IEEE Transactions on computers}, vol.~62, no.~11, pp. 2183--2195, 2012.

\bibitem{tehranipoor2010survey}
M.~Tehranipoor and F.~Koushanfar, ``A survey of hardware trojan taxonomy and detection,'' \emph{IEEE design \& test of computers}, vol.~27, no.~1, pp. 10--25, 2010.

\bibitem{gao2018watching}
Y.~Gao, B.~Li, W.~Wang, W.~Xu, C.~Zhou, and Z.~Jin, ``{Watching and Safeguarding your 3D Printer: Online Process Monitoring against Cyber-Physical Attacks},'' \emph{Proceedings of the ACM on Interactive, Mobile, Wearable and Ubiquitous Technologies}, vol.~2, no.~3, pp. 1--27, 2018.

\bibitem{stanczak2021vibration}
A.~Sta\'nczak, I.~Kubiak, A.~Przybysz, and A.~Witenberg, ``The possibility to recreate shapes based on printer vibration in additive printing,'' \emph{Applied Sciences}, vol.~11, 2021.

\bibitem{jamarani2024practitioner}
A.~Jamarani, Y.~Tu, and X.~Hei, ``Practitioner paper: Decoding intellectual property: Acoustic and magnetic side-channel attack on a 3d printer,'' in \emph{International Conference on Security and Privacy in Cyber-Physical Systems and Smart Vehicles}.\hskip 1em plus 0.5em minus 0.4em\relax Springer, 2024, pp. 54--74.

\bibitem{chattopadhyay2025one}
T.~Chattopadhyay, F.~Ceschin, M.~E. Garza, D.~Zyunkin, A.~Chhotaray, A.~P. Stebner, S.~Zonouz, and R.~Beyah, ``One video to steal them all: 3d-printing ip theft through optical side-channels,'' \emph{arXiv preprint arXiv:2506.21897}, 2025.

\bibitem{batina2019online}
L.~Batina, {\L}.~Chmielewski, L.~Papachristodoulou, P.~Schwabe, and M.~Tunstall, ``Online template attacks,'' \emph{Journal of Cryptographic Engineering}, vol.~9, no.~1, pp. 21--36, 2019.

\bibitem{zinner2022spooky}
T.~Zinner, G.~Parker, N.~Shamsaei, W.~King, and M.~Yampolskiy, ``{Spooky Manufacturing: Probabilistic Sabotage Attack in Metal AM using Shielding Gas Flow Control},'' in \emph{Proceedings of the 2022 ACM CCS Workshop on Additive Manufacturing (3D Printing) Security}, 2022, pp. 15--24.

\bibitem{vander2004laboratory}
G.~Vander~Voort, ``Laboratory safety in metallography,'' 2004.

\bibitem{osha2014after}
O.~R.~N. Release, ``After explosion, us department of labor's osha cites 3-d printing firm for exposing workers to combustible metal powder, electrical hazards powderpart inc. faces \$64,400 in penalties,'' 2014.

\bibitem{flukei310s}
{Fluke Corporation}, ``Fluke i310s ac/dc current clamp,'' \url{https://www.fluke.com/en-us/product/accessories/current-clamps/fluke-i310s}, 2025, accessed: 2025-05-20.

\bibitem{NIUSB6363}
{National Instruments}, ``{USB-6363 Multifunction I/O Device (Part No. 782258-01)},'' \url{https://www.newark.com/ni/782258-01/usb-6363-multifunction-i-o-device/dp/14AJ5493}, 2025, accessed: May 20, 2025.

\bibitem{NIflexlogger}
------, ``{NI FlexLogger},'' \url{https://www.ni.com/de-de/shop/product/flexlogger.html}, 2025, accessed: May 21, 2025.

\bibitem{NIdiadem}
------, ``{NI DIAdem},'' \url{https://www.ni.com/de-de/shop/product/diadem.html}, 2025, accessed: May 21, 2025.

\bibitem{nyquist1928certain}
H.~Nyquist, ``Certain topics in telegraph transmission theory,'' \emph{Transactions of the American Institute of Electrical Engineers}, vol.~47, no.~2, pp. 617--644, 1928.

\bibitem{shannon1948mathematical}
\BIBentryALTinterwordspacing
C.~E. Shannon, ``A mathematical theory of communication,'' \emph{Bell System Technical Journal}, vol.~27, no.~3, pp. 379--423, 1948. [Online]. Available: \url{https://ieeexplore.ieee.org/document/6773024}
\BIBentrySTDinterwordspacing

\bibitem{ASTME8}
\BIBentryALTinterwordspacing
{ASTM International}, ``{ASTM E8/E8M-22: Standard Test Methods for Tension Testing of Metallic Materials},'' ASTM International, Tech. Rep., 2022, accessed: May 21, 2025. [Online]. Available: \url{https://www.astm.org/Standards/E8}
\BIBentrySTDinterwordspacing

\bibitem{fitzgibbon1996direct}
A.~W. Fitzgibbon, M.~Pilu, and R.~B. Fisher, ``Direct least squares fitting of ellipses,'' in \emph{Proceedings of 13th international conference on pattern recognition}, vol.~1.\hskip 1em plus 0.5em minus 0.4em\relax IEEE, 1996, pp. 253--257.

\bibitem{todd2016minimum}
M.~J. Todd, \emph{Minimum-volume ellipsoids: Theory and algorithms}.\hskip 1em plus 0.5em minus 0.4em\relax SIAM, 2016.

\bibitem{hartley2003multiple}
R.~Hartley and A.~Zisserman, \emph{Multiple view geometry in computer vision}.\hskip 1em plus 0.5em minus 0.4em\relax Cambridge university press, 2003.

\bibitem{seber2003linear}
G.~A. Seber and A.~J. Lee, \emph{Linear regression analysis}.\hskip 1em plus 0.5em minus 0.4em\relax John Wiley \& Sons, 2003.

\bibitem{yampolskiy2018security}
M.~Yampolskiy, W.~E. King, J.~Gatlin, S.~Belikovetsky, A.~Brown, A.~Skjellum, and Y.~Elovici, ``{Security of Additive Manufacturing: Attack Taxonomy and Survey},'' \emph{Additive Manufacturing}, vol.~21, pp. 431--457, 2018.

\bibitem{belikovetsky2017dr0wned}
\BIBentryALTinterwordspacing
S.~Belikovetsky, M.~Yampolskiy, J.~Toh, J.~Gatlin, and Y.~Elovici, ``{dr0wned {\textendash} Cyber-Physical Attack with Additive Manufacturing},'' in \emph{11th {USENIX} Workshop on Offensive Technologies ({WOOT} 17)}.\hskip 1em plus 0.5em minus 0.4em\relax Vancouver, BC: {USENIX} Association, 2017, p.~16. [Online]. Available: \url{https://www.usenix.org/conference/woot17/workshop-program/presentation/belikovetsky}
\BIBentrySTDinterwordspacing

\bibitem{sturm2014cyber}
L.~Sturm, C.~Williams, J.~Camelio, J.~White, and R.~Parker, ``{Cyber-Physical Vulnerabilities in Additive Manufacturing Systems},'' \emph{Context}, vol.~7, p.~8, 2014.

\bibitem{zeltmann2016manufacturing}
S.~E. Zeltmann, N.~Gupta, N.~G. Tsoutsos, M.~Maniatakos, J.~Rajendran, and R.~Karri, ``{Manufacturing and Security Challenges in 3D Printing},'' \emph{JOM}, pp. 1--10, 2016.

\bibitem{graves2021sabotaging}
L.~Graves, W.~King, P.~Carrion, S.~Shao, N.~Shamsaei, and M.~Yampolskiy, ``{Sabotaging Metal Additive Manufacturing: Powder Delivery System Manipulation and Material-Dependent Effects},'' \emph{Additive Manufacturing}, vol.~46, pp. 1020--1029, 2021.

\bibitem{moore2017implications}
S.~B. Moore, W.~B. Glisson, and M.~Yampolskiy, ``{Implications of Malicious 3D Printer Firmware},'' in \emph{Proceedings of the 50th Hawaii International Conference on System Sciences}, 2017, pp. 6089--6098.

\bibitem{yampolskiy2015security}
M.~Yampolskiy, L.~Schutzle, U.~Vaidya, and A.~Yasinsac, ``{Security Challenges of Additive Manufacturing with Metals and Alloys},'' in \emph{Critical Infrastructure Protection IX}.\hskip 1em plus 0.5em minus 0.4em\relax Springer, 2015, pp. 169--183.

\bibitem{albakri2015non}
M.~Albakri, L.~Sturm, C.~B. Williams, and P.~Tarazaga, ``{Non-Destructive Evaluation of Additively Manufactured Parts via Impedance-Based Monitoring},'' in \emph{Solid Freeform Fabrication Symposium}, 2015, pp. 1475--1490.

\bibitem{chhetri2016kcad}
S.~R. Chhetri, A.~Canedo, and M.~A. Al~Faruque, ``{KCAD: Kinetic Cyber-Attack Detection Method for Cyber-Physical Additive Manufacturing Systems},'' in \emph{Proceedings of the 35th International Conference on Computer-Aided Design}.\hskip 1em plus 0.5em minus 0.4em\relax ACM, 2016, p.~74.

\bibitem{belikovetsky2019digital}
S.~Belikovetsky, Y.~A. Solewicz, M.~Yampolskiy, J.~Toh, and Y.~Elovici, ``{Digital Audio Signature for 3D Printing Integrity},'' \emph{IEEE Transactions on Information Forensics and Security}, vol.~14, no.~5, pp. 1127--1141, 2019.

\bibitem{bayens2017see}
\BIBentryALTinterwordspacing
C.~Bayens, T.~Le, L.~Garcia, R.~Beyah, M.~Javanmard, and S.~Zonouz, ``{See No Evil, Hear No Evil, Feel No Evil, Print No Evil? Malicious Fill Patterns Detection in Additive Manufacturing},'' in \emph{26th USENIX Security Symposium (USENIX Security 17)}.\hskip 1em plus 0.5em minus 0.4em\relax Vancouver, BC: USENIX Association, Aug. 2017, pp. 1181--1198. [Online]. Available: \url{https://www.usenix.org/conference/usenixsecurity17/technical-sessions/presentation/bayens}
\BIBentrySTDinterwordspacing

\bibitem{gatlin2019detecting}
J.~Gatlin, S.~Belikovetsky, S.~B. Moore, Y.~Solewicz, Y.~Elovici, and M.~Yampolskiy, ``{Detecting Sabotage Attacks in Additive Manufacturing using Actuator Power Signatures},'' \emph{IEEE Access}, vol.~7, pp. 133\,421--133\,432, 2019.

\bibitem{tsoutsos2017secure}
N.~G. Tsoutsos, H.~Gamil, and M.~Maniatakos, ``{Secure 3D Printing: Reconstructing and Validating Solid Geometries using Toolpath Reverse Engineering},'' in \emph{Proceedings of the 3rd ACM Workshop on Cyber-Physical System Security}.\hskip 1em plus 0.5em minus 0.4em\relax ACM, 2017, pp. 15--20.

\bibitem{anonymous_author_s_2025_16753249}
\BIBentryALTinterwordspacing
A.~Author(s), ``{Data Set: Power Side-Channel-based Design Reconstruction on PBF 3D Printer},'' Aug. 2025. [Online]. Available: \url{https://doi.org/10.5281/zenodo.16753249}
\BIBentrySTDinterwordspacing

\end{thebibliography}
